\documentstyle[psfig]{mn2e}
\title[Unusual high-redshift radio BAL quasar 1624+3758]{Unusual 
high-redshift radio
BAL quasar 1624+3758}
\author[C.R. Benn et al]{C.R. Benn,$^{1}$
\thanks{Email: crb@ing.iac.es}
R. Carballo,$^2$
J. Holt,$^3$
M. Vigotti,$^4$
J.I. Gonz\'{a}lez-Serrano,$^5$
K.-H. Mack$^4$,
\newauthor
R.A. Perley$^6$ \\
$^1$Isaac Newton Group, Apartado 321, E-38700 Santa Cruz de La Palma, Spain\\
$^2$Departamento de Matem\'atica Aplicada y C. Computaci\'on, 
Universidad de Cantabria,
E-39005 Santander, Spain\\
$^3$Department of Physics \& Astronomy, University of Sheffield, Hicks Building,
Sheffield S3 7RH, UK \\
$^4$Istituto di Radioastronomia, CNR, via Gobetti 101, 
I-40129 Bologna, Italy\\
$^5$Instituto de F\'\i sica de Cantabria (CSIC-UC), Facultad de Ciencias,
E-39005 Santander, Spain\\
$^6$National Radio Astronomy Observatory, PO Box 0, Socorro,
NM 87801, USA \\
}
\begin{document}
\maketitle

\begin{abstract}
We present observations of the most radio-luminous BAL quasar
known, 1624+3758, at redshift $z$ = 3.377.
The quasar has several unusual properties:
(1) The FeII UV191 1787-\AA\ emission line is very prominent.
(2) The BAL trough (BALnicity index 2990 kms$^{-1}$)
is detached by 21000 kms$^{-1}$ and extends to 
velocity $v$ =
-29000 kms$^{-1}$.
There are additional intrinsic absorbers at
-1900 and -2800 kms$^{-1}$.
(3) The radio rotation measure of the quasar, 
18350 rad m$^{-2}$, is the second-highest known.
\\
The radio luminosity is $P_{1.4GHz}$ = 
  4.3$\times$10$^{27}$
WHz$^{-1}$ ($H_0$ = 50 kms$^{-1}$Mpc$^{-1}$, $q_0$ = 0.5),
radio loudness $R^*$ = 260.
The radio source is compact ($\sol$ 2.8 kpc) and the radio spectrum is
GHz-peaked, consistent with it being relatively young.
The width of the CIV emission line, in conjunction with
the total optical luminosity, implies black-hole mass $M_{BH} 
\sim$ 10$^9$ $M_\odot$, $L/L_{Eddington} \approx$ 2.
The high Eddington ratio, and the radio-loudness,
place this quasar in one corner of Boroson's (2002) 2-component
scheme for the classification of AGN, 
implying a very high accretion rate, and this may account for
some of the unusual observed properties.
\\
The $v$ = -1900 kms$^{-1}$ absorber is a possible Lyman-limit system, with
$N$(HI) = 4$\times$ 10$^{18}$ cm$^{-2}$, and covering factor 0.7.
\\
A complex mini-BAL absorber at
$v$ = -2200 -- -3400 kms$^{-1}$ is detected
in each of CIV, NV and OVI.
The blue and red
components of the CIV doublet happen to be unblended, allowing
both the covering factor and optical depth to be determined
as a function of velocity.
Variation of covering factor with velocity
dominates the form of the mini-BAL, with the absorption 
being saturated (e$^{-\tau} \approx$ 0) over most of the
velocity range.
The velocity dependence of the covering factor, 
and the large velocity width, 
imply that the mini-BAL is intrinsic to the quasar.
There is some evidence of line-locking between 
velocity components in the CIV mini-BAL, 
suggesting that radiation pressure plays a role
in accelerating the outflow.

\end{abstract}


\section{Introduction}
In 10 -- 20\% of optically-selected quasars, 
broad absorption lines (BALs) are seen
in the blue wings of the UV resonance emission lines (e.g. CIV), 
due to gas with outflow velocities up to $\sim$ 0.2 c
(Hewett \& Foltz 2003).
The absorption troughs can be highly structured,
but are smooth compared with thermal line widths.
$\sim$ 20\% of BALs are detached from the corresponding emission line by 
several thousand kms$^{-1}$ (see Korista et al 1993 for examples).
The blue and red edges of the BAL absorption trough are often 
relatively abrupt, spanning
$\sim$ 100s kms$^{-1}$.
These distinctive features would be hard to reconcile with absorption
by individual clouds, but are consistent with the line of sight to 
a BAL quasar
intersecting an outflow which is not entirely
radial, e.g. an outflow which initially emerges perpendicular
to the accretion disk, and is then accelerated radially 
(Murray et al 1995, Elvis 2000).
NV BALs often absorb part of the Ly$\alpha$ emission line,
so the BAL region must typically lie outside at least some of
the broad emission-line 
region (BLR), i.e. $>$ 0.1 pc from the quasar nucleus.
BALs are generally saturated (optical depth $\sim$ few)
but non-black, implying partial covering of the nuclear regions
(or infilling of the absorption troughs by scattered light).
This means that column densities cannot be measured directly from apparent
absorption depths (in the past this has led to incorrect inference of
super-solar metallicities).

Formally, a BAL quasar is one with 
BALnicity index ($BI$, Weymann et al 1991) greater than zero.
$BI$ is defined as the equivalent width
of the CIV absorption, integrated over any contiguous region of the
spectrum 3000 - 25000 kms$^{-1}$ 
blueward of the quasar velocity, spanning at least 2000 kms$^{-1}$, with
continuum intensity $<$ 0.9
that of the assumed unabsorbed  continuum.
Shallower BAL-like features are seen in a larger fraction
of quasars
(Reichard et al 2004), suggesting that many quasars have similar
outflows.
Absorbers similar to quasar BALs are seen in Seyfert 1
galaxies, albeit with lower outflow velocities, typically $<$ 
few hundred kms$^{-1}$
(see contributions in Crenshaw, Kraemer \& George 2002).

The 
most prominent BALs are due to 
high-ionisation species, particularly Li-like ions
with one electron in the outer orbit: CIV 1549 \AA, SiIV 1400 \AA, 
NV 1240 \AA.
Quasars whose absorption is dominated by these
are known as high-ionisation BALs (HiBALs).
$\sim$ 15\% of BAL quasars also show absorption by
lower-ionisation species, 
such as MgII 2798 \AA\ and AlIII 1858 \AA, and are known as LoBALs.
FeLoBALs are a small subset of the LoBALs showing 
absorption by FeII and FeIII.

With few exceptions, no changes have been observed 
in the velocity structure of BALs  on
timescales $\sim$ 10 years.
The intensity of the absorption
does vary, probably due to changes in covering factor,
which suggests that the absorbers are intrinsic to the quasar.

BAL quasars are typically weak in soft X-rays, probably because
the X-ray emission is absorbed.
The relationship between UV and X-ray absorbers was discussed 
by Blustin et al (2004).

Several useful catalogues of BAL quasars exist.
Korista et al (1993) presented a sample of 72 CIV BALs.
Becker et al (2000, 2001) found 43 BALs in 
3300 deg$^2$ of the FIRST Bright Quasar Survey
($S_{1.4GHz} >$ 1 mJy).
Large samples of quasars are now becoming available from the
Sloan Digital Sky Survey (SDSS, York et al 2000).
Reichard et al (2003a) found 224 BAL quasars in the SDSS Early
Data Release quasar catalogue.
This sample includes the 116 BALs found
by Tolea, Krolik \& Tsvetanov (2002), and it overlaps with that of
Menou et al (2001), who sought identifications of SDSS quasars 
with FIRST sources in 290 deg$^2$ (14 radio BALs).
A catalogue of 23 unusual BALs found in SDSS (mainly LoBALs) was presented
by Hall et al (2002).

Hypotheses about the nature of BAL quasars differ mainly in
the emphasis placed on the role of orientation.
On the one hand, BALs may be present in all quasars but 
are intercepted by only $\sim$
10 -- 20\% of the lines of sight to the quasar, e.g. lines of sight
skimming the edge of the accretion disk or torus (Weymann et al 1991,
Elvis 2000).
Alternatively, BALs may arise in a physically
distinct population of quasars,
e.g. newborn quasars shedding their cocoons
of gas and dust, or quasars with unusually massive black holes,
or with unusually high accretion rates (Briggs, Turnshek \& Wolfe 1984,
Boroson \& Meyers 1992).
Amongst optically-selected quasars, 
evidence has accumulated to favour the orientation hypothesis.
E.g. in most respects, apart from the BAL itself, 
BAL quasars appear similar to normal quasars
(Weymann et al 1991).
The small differences from non-BAL quasars, e.g. slightly redder continua
(Reichard et al 2003b), and higher polarisation, could also be
a consequence of a preferred viewing angle.
The sub-mm properties of BALs are similar to those of non-BALs
(Willott, Rawlings \& Grimes 2003, Lewis, Chapman \& Kuncic 2003),
implying similar dust properties.  This is consistent with the
orientation hypothesis, but difficult to reconcile with BAL quasars
being an evolutionary stage associated with a large dust mass.

Until recently, very few radio-loud BAL quasars were known.
This changed with the advent of the FIRST Bright Quasar Survey (FBQS,
Becker et al 2001), but few BALs are known with
log $R^* >$ 2 (radio-loudness $R^*$ = $S_{5GHz}/S_{2500A}$,
Stocke et al 1992).  
Becker et al (2001) estimated that BALs are four times less common 
amongst quasars with log $R^* >$ 2 
than amongst quasars with log $R^* <$ 1.
Hewett \& Foltz (2003) note that optically-bright BAL quasars are half as
likely as non-BALs to have $S_{1.4GHz} >$ 1 mJy.
The dependence of BAL fraction on $R^*$ may reflect 
the higher ratio of X-ray to UV luminosity in radio-louder
objects, which could 
over-ionise the gas, reducing the velocity to which 
line-driven winds can be accelerated (Murray et al 1995).
Becker et al (2000) found that
radio-selected BAL quasars have a range of spectral indices, which
suggests a wide range of orientations, contrary to
the favoured interpretation for optically-selected
quasars.
Radio-loud BALs tend to be compact in the radio, similar to 
GPS (GHz-peaked spectrum) or CSS (compact steep-spectrum)
sources, and GPS/CSS sources are thought to 
be the young counterparts
of powerful large-scale radio sources (O'Dea 1998).
This supports the alternative hypothesis that BALs represent an early
phase in the life of quasars
(Gregg et al 2000).

In Boroson's (2002) scheme for the classification of AGN, 
based on a principal-component analysis of 
AGN properties, the different observed types
correspond to different combinations of $L/L_{Eddington}$ 
(luminosity as a fraction of Eddington luminosity)
and $dM/dt$ (the accretion rate).  BAL quasars occupy one corner of
this space, with $L/L_{Eddington} \sim$ 1, similar to narrow-line Sy1
galaxies, but with a much higher accretion rate.
The BAL quasar 
accretion rates are similar to those of radio-loud quasars, but
with larger $L/L_{Eddington}$ (and lower-mass black holes).
In Boroson's scheme,
the rare radio-loud BAL quasars may be objects with
extremely high accretion rates.

Lamy \& Hutsemekers (2004) carried out a principal-component analysis
of 139 BAL quasars with good-quality spectra and/or polarisation
measurements.
They found that most of the variation is contained in two principal
components.
The first is dominated by a correlation between $BI$ and the strength
of the FeII emission, and may be driven by the accretion rate.
The second is due to the fact that BALs with PCyg profiles (i.e. 
absorption just blueward of the emission line), 
are more polarised than those with detached BALs.
Detachment is thought to correlate with orientation,
with the more detached BALs being seen if the angle of the line of sight 
to the disk is larger.

Hewett \& Foltz (2003) found no evidence that 
the fraction of quasars with BALs varies with
redshift for $z <$ 3.
However, at higher redshift, the fraction may rise.
Maiolino et al (2004) recently found that of 8 $z >$ 4.9 quasars
observed, 4 showed strong BALs, with 2 of these having unusually 
high BALnicity
index, and two being LoBALs (which are rare at low redshift).
These results suggest that BALs are 
more common at high redshift, perhaps because of a higher accretion rate,
which might affect the solid angle 
subtended at the quasar by the BAL flow, and thus
the fraction of quasars observed to have BALs.

In short, the role of orientation in BALs is still not clear,
and it's likely that detailed measurements of physical conditions
within the outflows are required to make further progress.
Studies of BAL outflows are also important: 
(1) for understanding accretion in AGN,
where the inflow (fuelling) and outflow rates may be related through the
need to shed angular momentum;
(2) as probes of chemical enrichment in the central regions of AGN;
(3) because the outflowing gas may contribute significantly
to the metallicity of the IGM;
(4) because the outflows may affect the subsequent
evolution of the host galaxy (Silk \& Rees 1998, Fabian 1999);
and
(5) because the physics of the outflowing gas is not understood. 
No self-consistent physical model yet exists for the acceleration
of the gas, or, if the filling factor is small (many small clouds),
for its confinement.
Possible mechanisms for the acceleration include radiation pressure,
pressure from cosmic rays or centrifugally-driven magnetic disk winds
(de Kool 1997).
One possible signature of radiation pressure is absorption-absorption
line-locking, and this has been observed in a few quasars (see 
Section 3.2.2).

\subsection{NALs and mini-BALs}
The blending of saturated features in BALs precludes
measurement of the column densities, which are required to
constrain the ionisation balance, the distance
of the absorber
from the quasar and the physics involved in accelerating the outflows.
However, some quasars show additional narrow absorption lines
(NALs) with velocity widths small enough (FWHM $\sol$ 300 km/s)
that multiplets of individual ions can be resolved, 
allowing the covering factor and true
optical depth to be determined independently (Arav et al 1999).
Some NALs are intrinsic to the quasar (AALs, associated absorption
lines, see e.g. Wise at al 2004) and
may be related to the BAL phenomenon.
Intermediate in FWHM are the rarer ($\sim$ 1\% of quasars, 
Hamann \& Sabra 2004) mini-BALs,
FWHM $\sol$ 2000 km s$^{-1}$.
The partial covering,
variability and
smooth absorption troughs indicate that mini-BALs are intrinsic
outflows like those seen in BALs, but with the advantage that
in some cases the covering factor and optical depth can be measured as
a function of velocity.
Mini-BALs are thus particularly useful for 
constraining physical conditions in the outflow.

Using mostly the HST, Keck and VLT, 
high-resolution spectra of several intrinsic NAL and mini-BAL
absorbers have been
obtained, including those in
quasars (full names abbreviated)
 0011+0055 (Hutsem\'{e}kers, Hall \& Brinkmann 2004),
0300+0048 (Hall et al 2003),
 0449-13 (Barlow et al 1997),
 08279+5255 (Srianand et al 2000),
 0946+301 (Arav et al 2001),
 1037-2703 (Srianand et al 2001),
1044+3656 (deKool et al 2001),
 1230+0115 (Ganguly et al 2003),
 1303+308 (Foltz et al 1987, Vilkoviskij \& Irwin 2001),
 1415+3408 (Churchill et al 1999),
 1511+091 (Srianand et al 2002),
 1603+3002 (Arav et al 1999),
 1605-0112 (Gupta et al 2003),
 2233-606 (Petitjean et al 1999),
2302+029 (Jannuzi et al 1996),
UM675 (Hamann et al 1997), in
6 quasars studied by D'Odorico et al (2004), and in
the Sy1 galaxy
 NGC 5548 (Arav et al 2002).
The earlier of these papers established the intrinsic nature of the absorbers,
and showed the importance, when measuring column densities,
of taking into account saturation and the 
limited covering factor.
These analyses also implied that the absorbers lie close to the quasar nucleus
(although there are few actual measurements of distance),
and might in some cases be identified with the X-ray warm absorbers.

Some of the quasars (0449-13, 0946+301, 1037-2703, 1303+308)
show changes with time of covering factor.  
Only 1303+308 has shown
any change in {\it velocity} 
(55 kms$^{-1}$ increase in velocity over 5 rest-frame
years).
In some (0011+0055, 08279+5255, 2233-606), 
the covering factor varies with ion,
perhaps because of inhomogeneous coverage (Hamann \& Sabra 2004).
In 1603+3002 the absorption covers the 
continuum, but not the broad-line region (BLR).
1415+3408 is unusual in that the covering factor in NV is close to 1, 
suggesting an
unusual viewing angle.
Several of these objects (see Section 3.2.2) 
show evidence of line-locking between individual velocity
components.

In summary, intrinsic NALs and mini-BALs
are excellent probes of the abundances and physical
conditions in outflows
close to the nuclei of quasars, with each
object providing a fresh perspective.

\subsection{Radio BAL quasar 1624+3758}
Here we present a HiBAL quasar with unusual optical and radio
properties.
Quasar 1624+3758 (POSS $E$ = 18.1, $O-E$ = 2.5) 
was identified during a search for high-redshift quasars at the
positions of FIRST radio sources,
 using the Isaac Newton Telescope (Benn et al 2002,
Holt et al 2004).
The radio source, $S_{1.4GHz}$ = 56 mJy, is at
           RA  16$^h$ 24$^m$      53.47$^s$, 
           Dec 37$^o$ 58$^\prime$ 06.7$^{\prime\prime}$ (J2000),
0.02$^s$ E, 0.0'' N of the POSS-I/APM (Automated Plate Measuring Machine,
Cambridge)
optical position.
1624+3758 is the most radio-luminous BAL quasar known (Fig. 1),
$P_{1.4GHz}$ = 4.3$\times$10$^{27}$ WHz$^{-1}$.

It is also highly luminous in the optical.
The $E$ mag implies $M_{AB}$(1450-\AA) 
$\approx$ -27.6, 
$L_{1450A} \sim$ 5$\times$10$^{24}$ WHz$^{-1}$,
and $\nu L_\nu \sim$  10$^{47}$ ergs$^{-1}$, corresponding to
total luminosity $\sim$ 10$^{47.6}$ ergs$^{-1}$,
using the bolometric correction of Warner, Hamann \& Dietrich (2004).

1624+3758 is detected in the 2MASS survey (Cutri et al 2003),
with $J$ = 16.9, $H$ = 16.3, $K$ = 15.6.
No X-ray detection is recorded in
the NED/IPAC extragalactic database (NED).

The radio loudness of the quasar (defined above), calculated from the
radio spectrum reported in Section 2.2, and the 2MASS J mag (rest-frame
$\approx$ 2800 \AA), is $R^*$ = 260
(assuming no extinction in the UV).
This meets the conventional definition
of radio-loud, log$_{} R^* >$ 1.
Few BAL quasars are known with log $R^* >$ 2.

\begin{figure}
\begin{tabular}{c}
\psfig{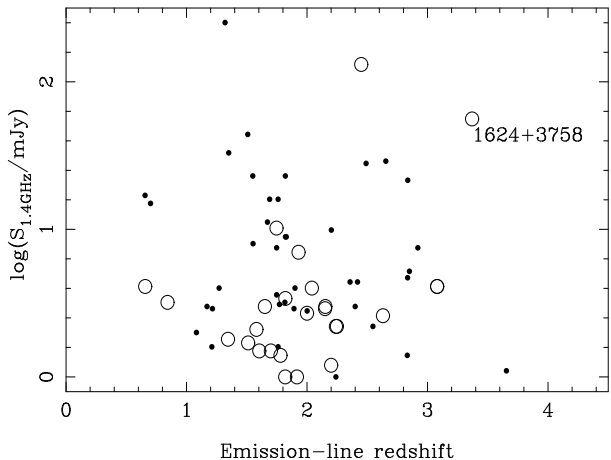}\\
\end{tabular}
\caption{Distribution of known radio BAL 
($BI >$ 0) quasars in 
emission-line redshift and 1.4-GHz flux
density (from the FIRST survey).
Circles indicate BALnicity index $BI >$ 2000 kms$^{-1}$.
Dots indicate 0 $< BI <$ 2000 kms$^{-1}$.
The following samples are plotted:
NVSS, Brotherton et al (1998, 5 quasars);
FBQS, Becker et al (2000, 2001, 33 quasars);
FIRST/SDSS, Menou et al (2001, 11 quasars);
SDSS, Reichard et al (2003a, 15 quasars); 
Brotherton et al (2002, two FRII BAL quasars, $S_{1.4GHz} >$
100 mJy); and
quasar 1624+3758 (labelled) reported in this paper.
1624+3758 is the most radio-luminous BAL quasar
identified to date.  Only the two FRII ($>$ 200 kpc) BAL quasars,
plotted near the top of the figure, 
have similar total 
radio luminosity.
Most of the remaining quasars are unresolved by the FIRST survey,
FWHM $<$ 5 arcsec.
}
\end{figure}


In this paper, we report
high-resolution ($R$ = 10000) optical spectroscopy of the quasar,
and radio observations.
In Section 2, we present the optical and radio observations.
The emission and absorption features are analysed in Section 3.
In Section 4 we discuss the nature of the quasar.
Our conclusions are summarised in Section 5.

For consistency with earlier papers, we use throughout
a cosmology with
$H_0$ = 50 kms$^{-1}$Mpc$^{-1}$, $q_0$ = 0.5, $\Lambda$ = 0.
Wavelengths are corrected to vacuum, heliocentric.

\vspace*{3mm}
\begin{figure}
\centering
\psfig{file=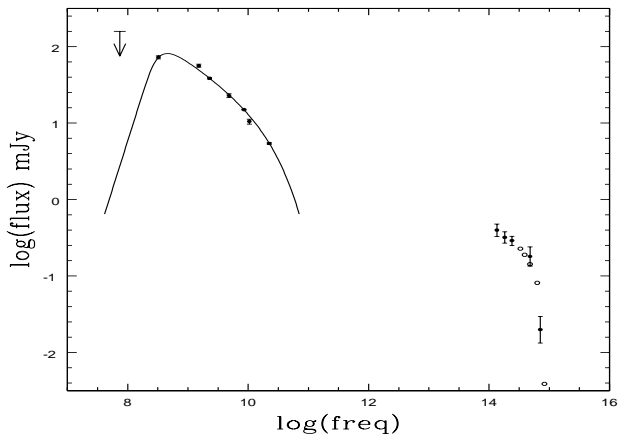,width=8.5cm,height=6cm,angle=0}
\caption{Spectral-energy distribution of BAL quasar 1624+3758.
The radio flux-density measurements are summarised in Table 2.
The higher-frequency measurements are from
the 2MASS survey ($K$, $H$, $J$ bands),
from the APM catalogues of objects on POSS-I ($E$, $O$ bands)
and from SDSS (no error bars plotted).
The radio spectrum turns over near 500 MHz, due to synchrotron
self-absorption at lower frequencies.
The solid curve is a fit by a synchrotron ageing model (Section 2.2).
}
\end{figure}

\section{Observations}
\subsection{Optical spectroscopy}
Spectra of 1624+3758 were obtained 
with ISIS, the dual-arm spectrograph of the 4.2-m William Herschel Telescope 
on La Palma, on
2003 June 18 (service observation) and on 2003 July 2.
The June spectrum covers the wavelength ranges 4200 - 5700
and 5700 - 7300 \AA, with
resolutions 2.3 and 1.8 \AA\ respectively.
The July spectrum covers the range 6200 - 7000 \AA, with resolution
0.8 \AA.
The observing details are summarised in Table 1.
All observations were carried out at the parallactic angle, in
photometric conditions, and with the moon below the horizon.

The data were reduced in the usual way,
using standard packages in IRAF for 
the bias subtraction, flat
fielding, cosmic-ray removal, wavelength calibration and intensity
calibration.
The rms errors in wavelength calibration,
determined by comparing the measured and published
(Osterbrock et al 1996) wavelengths of night-sky emission lines, are given in 
column 11 of Table 1 (the value for the June 18 blue-arm observation
is an upper limit, since only a few faint sky lines were detected).
The spectral resolution (column 10)
was measured from the widths of the 
sky lines.

Inter-comparison of the standard-star spectra
suggests that the accuracy of the intensity calibration
is $\sim$ 5\%.

A low-resolution spectrum of the quasar (FWHM $\approx$ 4 \AA\ in the red) 
was obtained in 2003 by 
the Sloan Digital Sky Survey (SDSS data release 3,
Abazajian et al 2004).  The SDSS apparent magnitudes are
$u$ = 22.4, $g$ = 19.1, $r$ = 18.5, $i$ = 18.2, $z$ = 18.0.

\begin{center}
\begin{table*}
\begin{minipage}{170mm}
\caption{Log of observations.}
\label{Table 1}
\begin{tabular}{lccllllcrcr}\\  \hline
\multicolumn{1}{c}{Date of}
& \multicolumn{1}{c}{Airmass}
& \multicolumn{1}{c}{Seeing} 
& \multicolumn{1}{c}{Dichroic} 
& \multicolumn{1}{c}{ISIS} 
& \multicolumn{1}{c}{Grating} 
& \multicolumn{1}{c}{CCD} 
& \multicolumn{1}{c}{Wavel. range}
& \multicolumn{1}{c}{Exposure}
& \multicolumn{1}{c}{Resolution} 
& \multicolumn{1}{c}{$\sigma_\lambda$} \\ 
observation
&
& \multicolumn{1}{c}{(arcsec)}  
& used
& arm
& 
&
& \multicolumn{1}{c}{(\AA~)} 
& \multicolumn{1}{c}{(s)} 
& \multicolumn{1}{c}{(\AA~)}
& \multicolumn{1}{c}{(\AA~)} \\ 
\multicolumn{1}{c}{(1)}
&\multicolumn{1}{c}{(2)} 
& \multicolumn{1}{c}{(3)}
& \multicolumn{1}{c}{(4)} 
& \multicolumn{1}{c}{(5)} 
& \multicolumn{1}{c}{(6)} 
& \multicolumn{1}{c}{(7)} 
& \multicolumn{1}{c}{(8)}
& \multicolumn{1}{c}{(9)}
& \multicolumn{1}{c}{(10)}
& \multicolumn{1}{c}{(11)} \\ 
\hline
2003 June 18 & 1.02 & 0.9 & 5700 \AA\
  & blue & R600B & EEV12 & 4200 -- 5700 & 2$\times$1800 & 2.3 & 0.15 \\
2003 June 18 & 1.02 & 0.9 & 5700 \AA\
  & red  & R600R & MAR2  & 5700 -- 7300 & 2$\times$1800 & 1.8 & 0.04 \\
2003 July 2  & 1.73 & 0.9 & clear
  & red  &R1200R & MAR2  & 6200 -- 7000 & 3$\times$1800 & 0.8 & 0.04 \\
\hline
\end{tabular}
\end{minipage}
\end{table*}
\end{center}


\subsection{Radio observations}
We observed 1624+3758 with the 100-m Effelsberg radio telescope 
at 4.85 and 10.45 GHz 
on 2003 Dec 18 and 2004 Jan 17 respectively.
The cross-scanning technique 
used is described by Vigotti et al (1999)
and references therein.  3C286 was used as a flux-density calibrator.

We also observed 1624+3758 with the VLA (Very Large Array) radio
telescope in C configuration, 
at 8.5 GHz (exposure time 20 mins) and 22.5 GHz (exposure time 40 mins), 
on 2004 Mar 24.
The nearby source 1613+342 was used as a phase calibrator.
Initial images of the source contained sufficient flux 
density to permit local 
self-calibration to remove residual phase errors.
3C286 was used as a flux-density calibrator.
The data were reduced with the IMAGR program 
in the AIPS package.

We also observed the quasar with the Westerbork Synthesis Radio Telescope
(WSRT) for 3 hours on 2004 Jul 1.
The 2.2-GHz flux density was measured, but only a 3-sigma upper limit
on the polarisation could be obtained, due to a combination of
radio interference and technical problems.

Our radio observations are summarised, with others from the literature,
in Table 2, and the spectral energy distribution is shown in Fig 2.
The spectrum is steep at high radio frequencies, $\alpha$ = -0.9
($S_\nu \propto \nu^\alpha$).

At 22.5 GHz, the source is resolved by the VLA, 
size 0.4 $\pm$ 0.1 arcsec. 
This implies a projected linear
size $\approx$ 2.8 kpc 
i.e. this is a Compact Steep Spectrum (CSS) source,
as are most radio BAL quasars (Becker et al 2000).
1624+3758 is one of the most distant CSS sources known.
The spectrum turns over at low frequency, 
probably due to synchrotron self-absorption.
The rest-frame turnover frequency $\approx$ 2 GHz
provides an independent estimate of size, 
$\sim$ 0.1 -- 1 kpc (using
the relationship given by O'Dea 1998).

The steep radio spectral index implies that the source is
lobe-dominated. 
The spectrum at frequencies higher than 1 GHz shows significant
curvature (it cannot be fitted with a simple power law), which we
attribute to ageing of the population of relativistic electrons
responsible for the synchrotron emission 
(Murgia et al 1999).
We fitted the observed spectrum with two popular ageing models:
the continuous-injection (CI) model  of Pacholzcyk (1970),
which assumes continuous replenishment within each resolution element
of particles and energy lost through synchrotron radiation;
and the model of Jaffe \& Perola (1973, JP),
which predicts spectral ageing assuming no replenishment.
Fitting the spectrum with these models yields the break frequency
$\nu_{br}$ at which the spectrum starts to turn down as a result of
synchrotron losses.
The particle ages are then given by:
\\
$\tau = 1610 \frac{B^{0.5}}{B^2+B^2_{CMB}} 
\frac{1}{\sqrt{\nu_{\rm br}(1+z)}}$ My
\\
(Murgia et al 1999),
where $B$ is
the source magnetic field in $\mu$G,
$B_{CMB}$ is the magnetic field strength corresponding to the
cosmic microwave background ($B_{CMB} = 3.25 (1+z)^2 \mu$G),
and $\nu_{br}$ is in GHz.
The JP model fits the data better than the CI model, as
is often the case for relic sources in which activity has ceased.
The fitted break frequencies $\nu_{br}$ are $7.6^{+1.7}_{-1.3}$ GHz for 
the CI model
and $38.0^{+4.8}_{-3.8}$ GHz for the JP model. 
If equipartition of energy between magnetic field $B$
and relativistic particles is assumed, then
(following Miley 1980) 
$B$ = 740 $\mu$G.
Even using the lower estimated $\nu_{br}$
(from the CI model), this yields a 
maximum particle age of $< 15000$ y, is
typical of synchrotron ages found 
for lobe-dominated CSS sources (e.g. Murgia et al
1999).

The radio emission in 
1624+3758 is  therefore of recent origin compared with the typical age 
of an evolved radio source,
$\sim 10^7$ y.

The position angle of the polarisation PA was measured at each 
of 4.85, 8.46, 10.45 and 22.46 GHz
(the source is depolarised at frequencies $\le$ 2.2 GHz).
From the variation with wavelength (Fig. 3), we determine
the rotation measure $RM$
($\Delta{\rm PA} = RM \cdot \Delta\lambda^2$) to be
960 $\pm$ 30 rad\,m$^{-2}$ which in the
rest frame of the source is   
a factor (1 + z)$^2$ higher, i.e. 18350 $\pm$ 570 rad\,m$^{-2}$.
This is the second-highest $RM$ known, after that of
quasar OQ172 (Kato et al 1987, O'Dea 1998), $RM$ = 22400 rad\,m$^{-2}$.
OQ172 is not a BAL quasar.
In a recently published compilation of pc-scale $RM$ in AGN, 
Zavala \& Taylor (2004) found a median $RM$ of 2000
${\rm rad\,m}^{-2}$ and no AGN with $RM >$ 10000 ${\rm rad\,m}^{-2}$.

\begin{figure}
\centering
\psfig{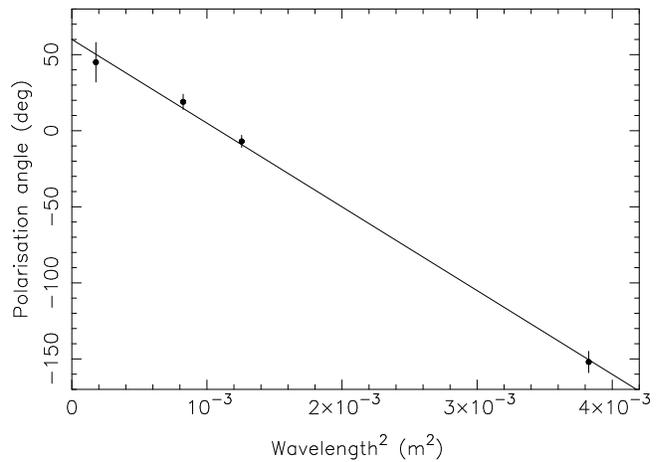}
\caption{Derivation of the rotation measure RM from the
radio observations.
The straight line has slope $RM$ = -960 rad m$^{-2}$.
This is the second-highest $RM$ known amongst AGN.
}
\end{figure}

Rotation measure is proportional to $n_e B_{||} l$, where $n_e$
is the electron density, 
$B_{||}$ is the magnetic field strength along the
line-of-sight and $l$ 
the effective path length along the line-of-sight,
so an unusually high value of $RM$ implies a high value of 
at least one of these parameters.

\begin{table*}
\begin{minipage}{170mm}
\begin{center}
\caption{Radio observations of 1624+3758}
\label{Table 1}
\begin{tabular}{lrrrrrrrrl}\\  \hline
Telescope & Freq. & Resolution & 
\multicolumn{2}{c}{Flux density} & 
\multicolumn{2}{c}{Polarisation} & 
\multicolumn{2}{c}{$PA$} & 
Survey \\
& GHz & arcsec & mJy & $\pm$ &\%&$\pm$&deg&$\pm$& \\
\hline
VLA (B) & 0.07\rlap{4} &80& $<$ 220& &&&&   & VLSS \\
WSRT& 0.32\rlap{5} &54& 72 & 5.3   &&&&& WENSS \\
VLA (D) & 1.40 &45& 55.6 & 1.7   &$<$ 2.2&&&& NVSS \\
VLA (B) & 1.40 &5& 56.4   & 1.5&&&&& FIRST \\
WSRT    & 2.27 & 36&38.5 & 0.8&$<$ 2.3&&&& this paper \\
Effelsberg &  4.85 & 120& 23.3 &1.1 &1.7&0.7&-152&7& this paper \\
VLA (C) &  8.46 &2\rlap{.4}& 15.0 & 0.0\rlap{9} &6.5&0.3&-7&4& this paper \\
Effelsberg & 10.45 & 270& 10.5 &0.8 &11.0&3.0&19&5& this paper \\
VLA (C) & 22.46 &0\rlap{.9}&  5.4 & 0.0\rlap{2} &11.3&1.5&45&13& this paper \\
\hline
\end{tabular}
\\
\end{center}
The flux densities are all on the scale of Baars et al (1977).
The surveys are: VLSS = VLA Low-frequency Sky Survey
(Kassim et al 2003);
WENSS = Westerbork Northern Sky Survey (Rengelink et al 1997);
NVSS = NRAO VLA Sky Survey (Condon et al 1998);
FIRST = Faint Images of the Radio Sky at Twenty-cm 
(White et al 1997).
$PA$ = position angle of polarisation.
\end{minipage}
\end{table*}

\section{Results of optical spectroscopy}
Fig. 4 shows the ISIS spectrum of 1624+3758 obtained 2003 Jun 18.
Figs. 5 and 6 show the spectrum at higher dispersion.
Emission and absorption features are listed in Tables 3 and 4
respectively.
Derived column densities are listed in Table 5.

We discuss the emission lines in Section 3.1.
The absorption features include:
a mini-BAL (defined as a BAL-like feature spanning 
$<$ 2000 kms$^{-1}$, Section 3.2, Figs. 8, 9, 10);
a possible associated (i.e. 
$v >$ -3000 kms$^{-1}$)
absorber with large HI column density
(Section 3.3);
a BAL (Section 3.4);
and 9 NALs (2 probably associated,  
7 intervening, 
Section 3.5).

The spectral index of the continuum between observed wavelengths 7000
and 9000 \AA\ (rest-frame 1600 - 2060 \AA) is $\alpha_\lambda$ = -0.7  
($f_\lambda \propto \lambda^\alpha_\lambda$), $\alpha_{\nu}$ = -1.3,
slightly redder than most HiBAL quasars (Reichard et al 2003b),
similar to the median for LoBAL quasars,
and suggesting dust reddening $E(B-V) \approx$ 0.07 mag relative
to non-BAL quasars, assuming extinction by SMC-type dust (Pei 1992).
(The quasar's much redder broad-band colour, $O-E$ = 2.5, 
is due to the drop in the continuum across Ly$\alpha$.)

\begin{figure*}
\centering
\psfig{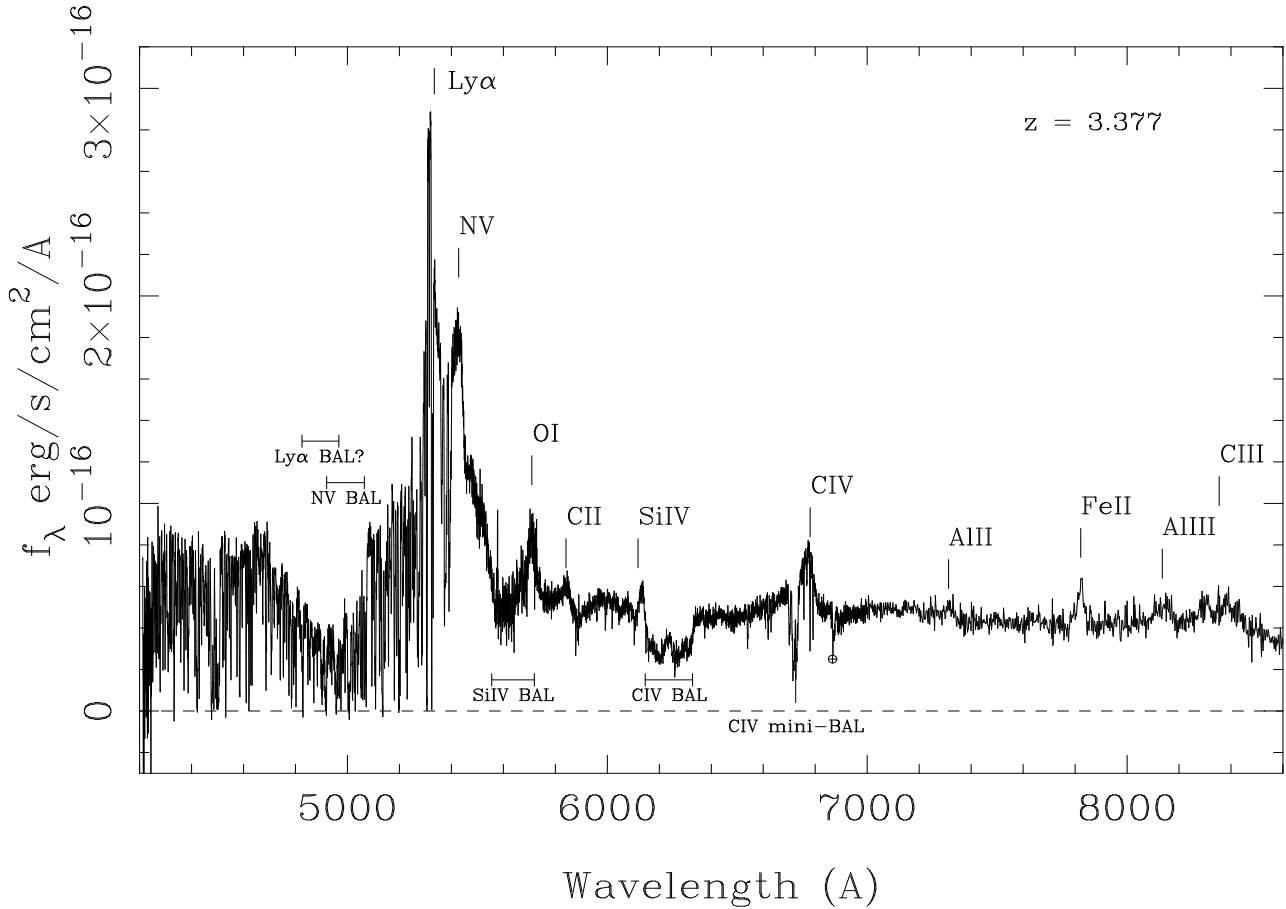}
\caption{Spectrum of BAL quasar 1624+3758 taken with the WHT ISIS spectrograph
2003 Jun 18.
The blue- and red-arm spectra are joined at 5730 \AA.
The spectrum redward of 7000 \AA\ is from the lower-resolution
observation of this quasar in SDSS data release 3.
The ticks above the emission lines indicate the wavelengths expected
for redshift 3.377, assuming the $\lambda_{lab}$ wavelengths given in
Table 3.
Horizontal bars indicates the CIV BAL outflow,
velocity -20700 -- -29300 kms$^{-1}$, and the expected wavelengths of 
absorption by
HI, NV and SiIV ions with a similar range of velocity.
The mini-BAL just blueward of the CIV emission line ranges in velocity
-2200 -- -3400 kms$^{-1}$.
It is also detected in NV ($\approx$ 5370 \AA, see Fig. 5) and in
OVI ($\approx$ 4490 \AA, see Fig. 5).
The feature at 6867 \AA\ is uncorrected telluric absorption
(B band).
}
\end{figure*}

\begin{figure*}
\centering
\psfig{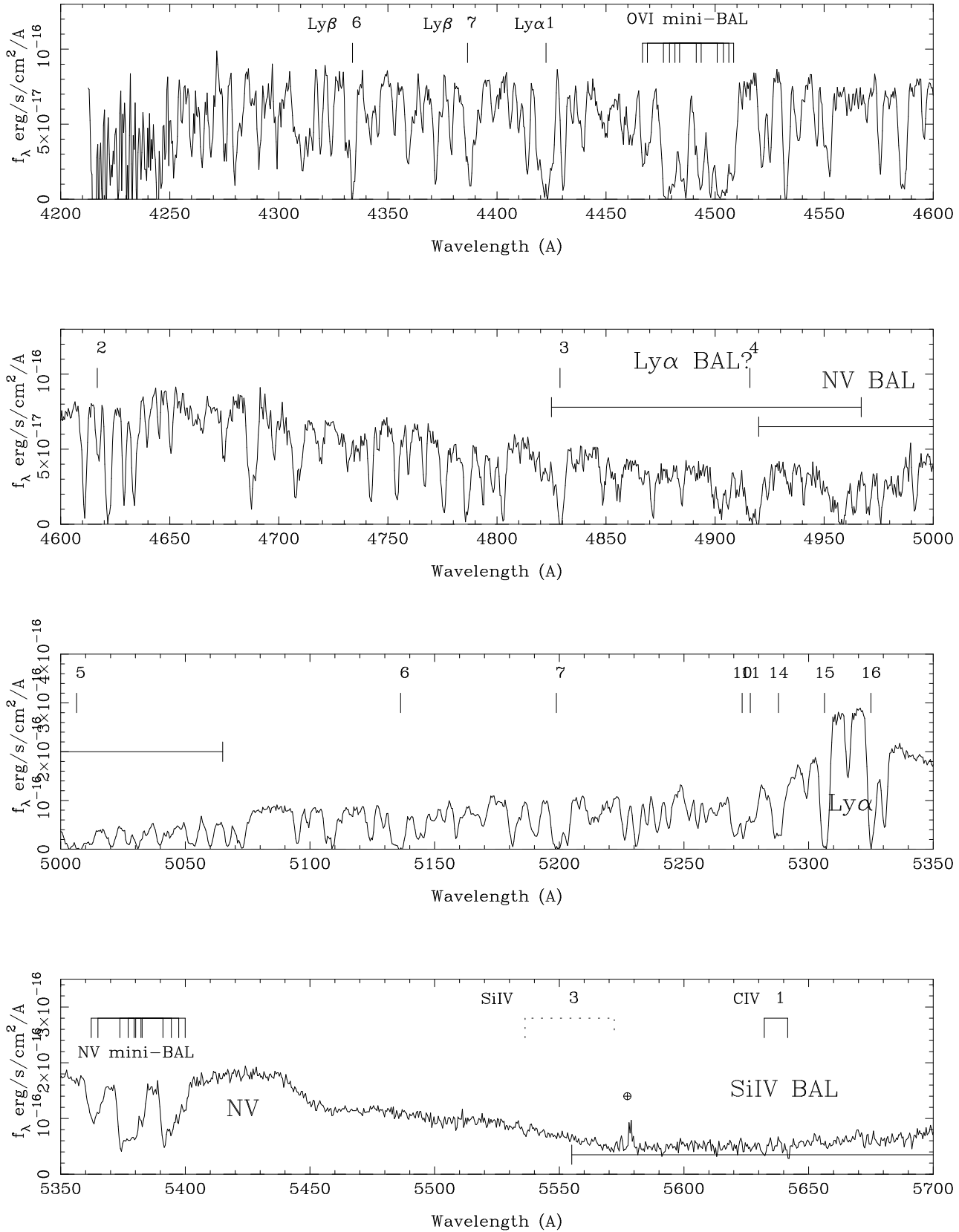}
\caption{WHT ISIS blue-arm spectrum of BAL quasar 1624+3758, 
taken 2003 Jun 18, plotted at higher dispersion than in Fig. 4.
The ticks indicate 
detected absorption features 
(Ly$\alpha$ unless otherwise labelled)
corresponding to the redshifts
listed in Table 4
(laboratory wavelengths given in Table 3).
The quasar Ly$_\alpha$ and NV emission lines are also labelled
(large font).
The long horizontal bars indicate the expected range of wavelengths of 
absorption by Ly$\alpha$, NV and SiIV ions at the same velocity as the CIV BAL.
A feature due to poor subtraction of the 5577-\AA\ airglow is marked.
}
\end{figure*}

\begin{figure*}
\centering
\psfig{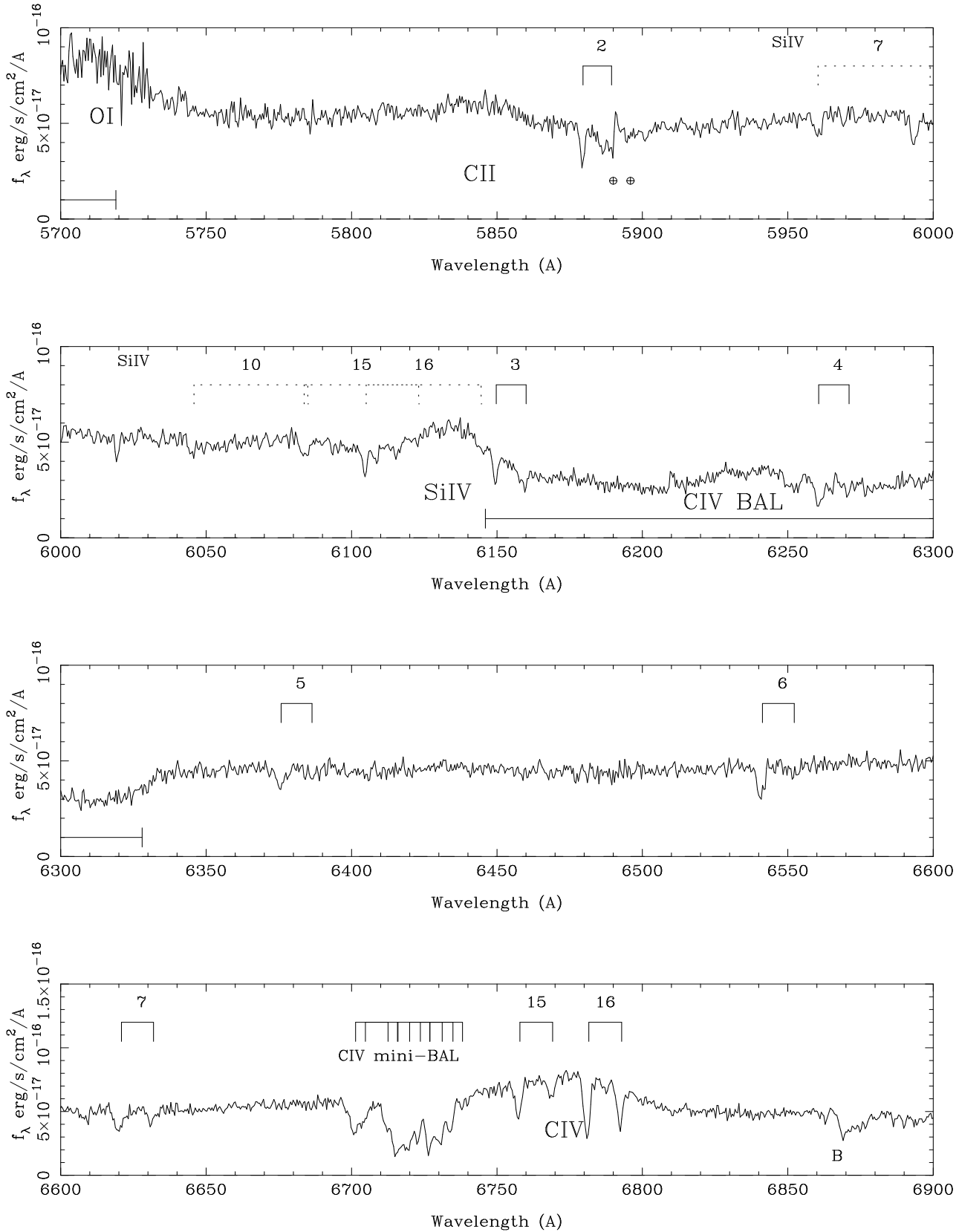}
\caption{WHT ISIS red-arm 
spectrum of quasar 1624+3758, taken 2003 Jun 18, plotted at higher dispersion
than in Fig. 4,
as in Fig. 5.  
The OI, CII, SiIV and CIV emission lines are labelled.
The ticks indicate CIV (solid lines) and SiIV (dotted lines)
absorption doublets.
The red component of the CIV doublet of absorber 2 may be confused by
the blue component of the NaD sky-line doublet (indicated).
'B' marks the uncorrected atmospheric absorption feature at
at 6867 \AA. 
The weak absorption at 5992 and 6018 \AA\ (also seen in the SDSS
spectrum) could be additional CIV 
absorbers, but in neither case is the second component of the 
doublet detected.
}
\end{figure*}

\subsection{Emission lines, and quasar redshift}
The centroid wavelengths 
of the Ly$\alpha$, NV, OI, CII, SiIV and CIV
emission lines are given in column 4 of Table 3.
Ly$\beta$ 1026 \AA, OVI 1035 \AA\ and 
HeII 1640 \AA\ are not detected in emission.

\begin{table*}
\begin{minipage}{170mm}
\begin{center}
\caption{Emission lines in the spectrum of 1624+3758}
\begin{tabular}{rrrrrrrrrl}
\hline
Line & $\lambda_{lab}$ &$\lambda_{sr}$ & $\lambda_{observed}$
& $\pm$ & $z_{qso}$ & $\pm$ 
& $v$ &$v_{sr}$&Notes\\
& \AA\ & \AA\ & \AA\ & \AA\ & & & kms$^{-1}$ & kms$^{-1}$& \\
(1) & (2) & (3) & (4) & (5) & (6) & (7) & (8) & (9) & (10) \\
\hline
Ly$_\alpha$  
     &1215.67 
 & 1214.97 & 5314.7&3  &3.3743  &0.0025  &-350&-180& \\
NV   &1238.82 / 1242.80
 & 1239.16 & 5420.2&2  &3.3741  &0.0016  &-430&-200&\\
OI   &1302/4/6 triplet 
 &1304.24 & 5713.6&5  &3.3808  &0.0038  &210& 260& confused by SIV BAL?\\
CII  &1334.53
 & 1334.53 & 5844.7&3  &3.3796  &0.0022  &150& 180&\\
SiIV &1393.75 / 1402.77
 & 1398.62 & 6131.6&3  &3.3840  &0.0021  &630& 480&  SiIV/OIV$^*$\\
CIV  &1548.20 / 1550.78
 & 1547.46 & 6771.2&3  &3.3757  &0.0019  &-320& -90&\\
\hline
\end{tabular}
\\
\end{center}
The columns give: 
(1) line; (2) laboratory vacuum wavelength; 
(3) representative wavelength tabulated by Tytler \& Fan (1992)
for obtaining $z_{qso}$;
(4, 5) observed wavelength and error;
(6, 7) implied quasar redshift and error; 
(8) velocity of line relative to the assumed
quasar $z$ = 3.377 (negative velocity
= blueshift);
(9) velocity of line 
relative to mean from Tytler \& Fan; (10) notes.\\
The weaker emission lines detected in the SDSS spectrum redward
of 7000 \AA\ have not been used for estimating the redshift.
\\
For convenience, we note here 
the laboratory vacuum wavelengths of the other lines
mentioned in this paper: Ly$\beta$ 1025.72; OVI 1031.93, 1037.62;
AlII 1670.79; FeII 1785/7/8; AlIII 1854.72 1862.79; CIII] 1908.73
\AA.
\\
$\ast$The doublet wavelengths are for SiIV.  The SiIV emission is usually
blended with an OIV multiplet at 1402 \AA.\\
\end{minipage}
\end{table*}

The high-ionisation lines NV and CIV are blueshifted a few 100 kms$^{-1}$
with respect
to the low-ionisation lines, as in most quasars (Gaskell 1982,
Tytler \& Fan 1992, Richards et al 2002b).
To obtain a system redshift $z_{sys}$ (usually
closer to that of the low-ionisation lines
than that of the high-ionisation lines),
we use the modified emission-line wavelengths $\lambda_{sr}$ 
tabulated by Tytler \& Fan.  These
take into account the velocity shifts, and approximate
$\lambda_{observed}/(1+z_{sys})$.
Tytler \& Fan do not give a modified wavelength for CII, but this is a 
low-ionisation line,
so we assume zero velocity shift relative
to the quasar, i.e. $\lambda_{sr}$ = $\lambda_{lab}$ =
1334.53 \AA.
The observed emission-line wavelengths
are consistent with a quasar redshift $z$ = 3.377 $\pm$ 0.003.
Column 8 of Table 3 gives the velocity of each line relative to 
this redshift.
Column 9 gives the velocity relative to the mean for that line
found by Tytler \& Fan (typical 
dispersions $\sim$ 200 kms$^{-1}$).
The SiIV line in 1624+3758 is significantly redshifted relative to 
the other lines, but is blended with an OIV emission line,
which might be strong in this quasar, given that the
OI 1304 \AA\ line is unusually strong.

The CIV emission line is blueshifted relative to the 
quasar by 320
kms$^{-1}$, similar to the shifts found by Tytler \& Fan for other quasars.
The line is markedly asymmetric at its base (and the NV emission
line has similar form, Fig. 7), consistent with the 
suggestion by
Richards et al (2002b) that the apparent blueshift of CIV lines
is due to the red wings of the lines being 
suppressed, perhaps in part due to 
dust obscuration of emission from
outflows on the far side of the nucleus (see their fig. 11).

The rest-frame FWHM of the CIV line (inferred from the blue half of the line)
is 2300 kms$^{-1}$.
In conjunction with the observed total luminosity (Section 1.2),
this implies (Kaspi et al 2000;
Warner, Hamann \& Dietrich 2004) a black-hole-mass
$M_{BH} \sim$  10$^9 M_\odot$.
This is similar to masses determined by Lacy et al (2001) for quasars 
of similar radio luminosity.
It implies a high Eddington ratio $L/L_{Eddington} \approx$ 2.0,
near the maximum found for quasars by Warner et al 
(2004).

The OI line is prominent, $EW \sim$ 4 \AA, compared with 1.7 \AA\ in
the SDSS composite quasar spectrum of Vanden Berk et al (2001),
and $\sim$ 1.8 and 5.0 \AA\ in the composite HiBAL and LoBAL spectra
of Reichard et al (2003a).
Bright OI emission is more common amongst those Reichard et al
BAL quasars with FIRST radio detections, $S_{1.4GHz} >$ 1 mJy
(6 out of 9, $z <$ 3) than amongst those without 
radio detection (7 out of 31, $z <$ 3).

The prominent emission line at observed wavelength 
7822.7 \AA, $EW$ = 3.8 \AA,
corresponding to rest-frame 1787.2 \AA\ (Fig. 4), is detected
in the SDSS composite spectrum,
and is probably the 
FeII UV191 triplet at 1785/1787/1788 \AA.  
In 1624+3758, this line is unusually strong, with rest-frame $EW$ =
3.8 \AA, compared to 0.3 \AA\ in the SDSS composite spectrum, 
and $\sim$ 0.4 and 1.7 \AA\ in the composite HiBAL and LoBAL spectra
of Reichard et al (2003a).
The FeII/CIV $EW$ ratio is also high $\sim$ 0.5, compared to 0.01 for the 
SDSS composite, 
and $\sim$ 0.02 and 0.1 for the Reichard et al HiBAL and LoBAL 
composites.

FeII 1787-\AA\ is
detected in the SDSS spectra of only 1 out of 40 randomly-selected
non-radio BALs (half $z <$ 3, half $z >$ 3) from the catalogue of Reichard
et al (2003a), and in only 1 out of 20 non-radio
LoBAL quasars from that catalogue.
However, it's detected in the spectra of 4 out of the 14 BALs from
that catalogue which have radio counterparts in the FIRST catalogue,
$S_{1.4GHz} >$ 1 mJy.
The FeII/CIV ratio in 1624+3758 is higher than in any of these objects.

The two relatively-narrow emission lines
at 1896.2 and 1915.1 \AA, superimposed on the 
broad (and weak) CIII] emission line (Fig. 4) are probably
the two bluer components of the FeIII
UV34 triplet at 1895.5/1914.1/1926.3 \AA.

These FeII and FeIII emission lines have also been observed in
BAL quasar H0335-336 (Hartig \& Baldwin 1986), and in
quasar 2226-3905 (Graham et al 1996).
Graham et al note that the FeII UV191 line is probably produced by
dielectric recombination from Fe$^{2+}$, i.e. it's not surprising 
to observe it in asociation with FeIII lines.

\subsection{Mini-BAL absorber, 3.3283 $< z <$ 3.3450}

\begin{figure}
\centering
\psfig{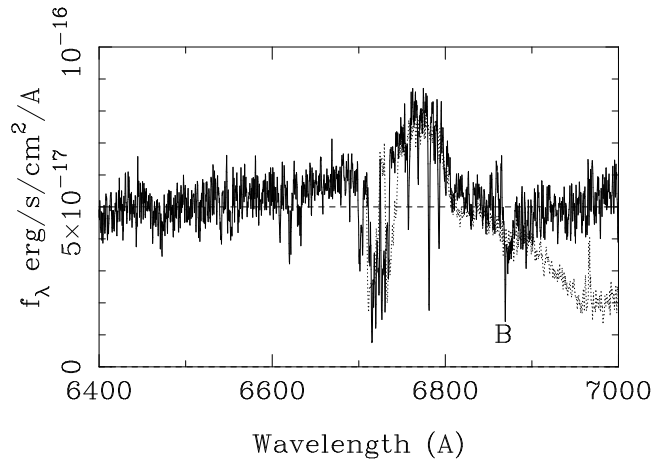}
\caption{The CIV emission line (6771 \AA) is 
asymmetric, with a broader wing to the blue than to the red.
The dashed horizontal line shows the approximate level of the
underlying continuum.  
The overplotted (dotted) curve shows the 
similar asymmetry in the core of the NV emission line.
}
\end{figure}

\begin{center}
\begin{figure*}
\begin{center}
\psfig{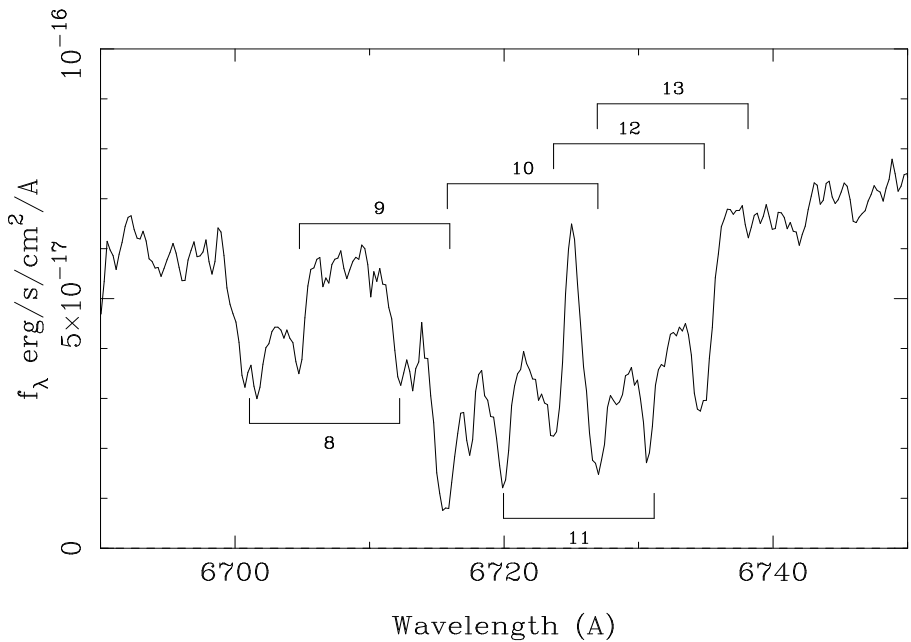}
\end{center}
\caption{The spectrum of the CIV mini-BAL, taken with ISIS 2003 Jul 2
(instrumental resolution 0.8 \AA, or 36 kms$^{-1}$).
The ticks indicate the expected observed-frame wavelengths
of the CIV doublet (1548.20, 1550.78 \AA), for each of the
redshifts identified in Section 3.2 (Table 4).
The expected wavelengths for component 13 (detected in NV only, Fig. 9) are
also indicated.
}
\end{figure*}
\end{center}

\begin{figure*}
\centering
\psfig{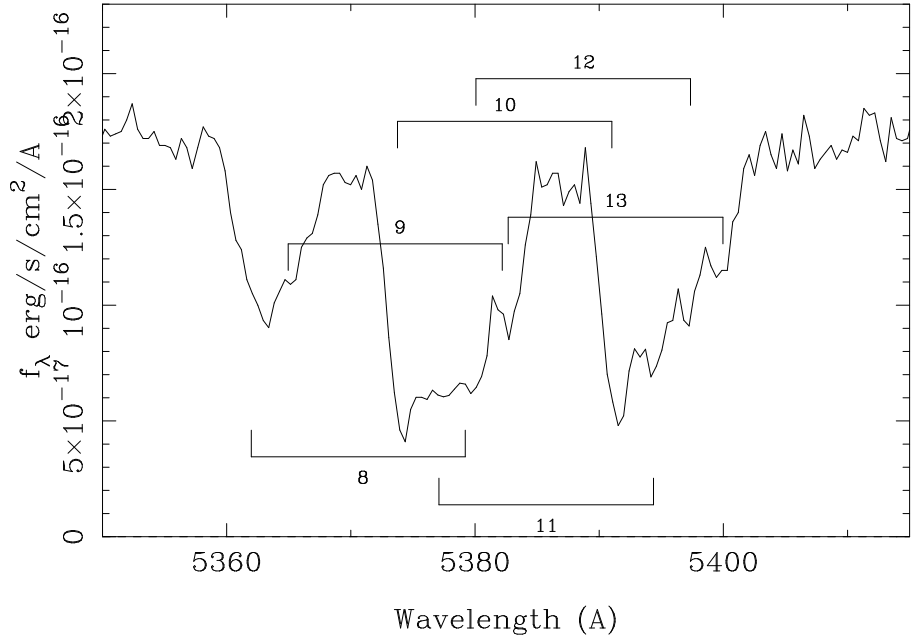}
\caption{The mini-BAL in NV from the 2003 Jun 18 spectrum
(the July spectrum does not include these wavelengths).
The ticks indicate the expected observed-frame wavelengths
of the NV doublet (1238.82, 1242.80 \AA), for each of the
redshifts in Fig. 8. 
Absorber 13 is detected only in NV.
The spectral resolution is a factor $\approx$ 3 poorer than
that of Fig. 8.
}
\end{figure*}

\begin{figure*}
\centering
\psfig{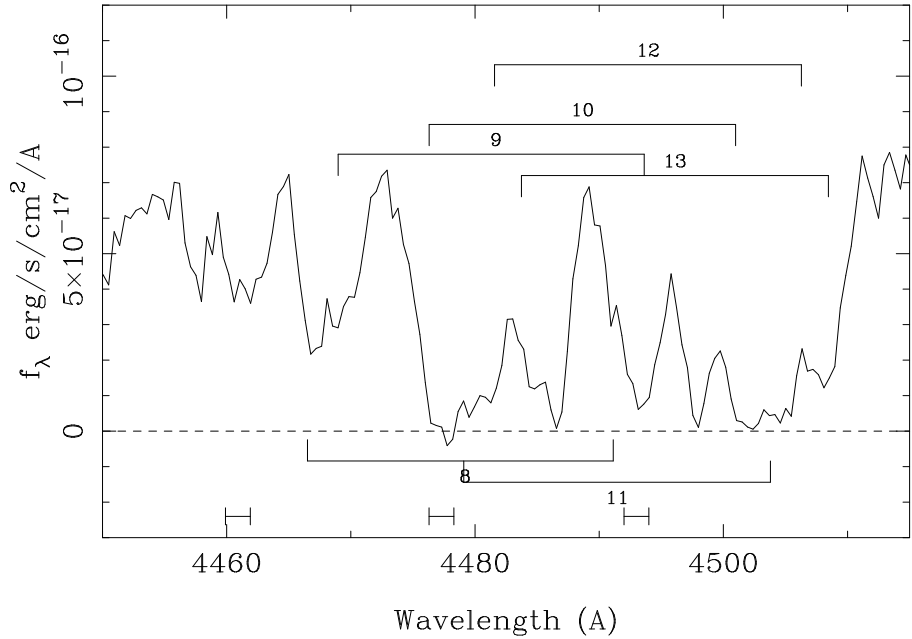}
\caption{The mini-BAL in OVI from the 2003 Jun 18 spectrum.
The ticks indicate the expected observed-frame wavelengths
of the OVI (1031.93, 1037.62 \AA) absorption, for each of the components
shown in Figs 8, 9.
This region of the spectrum falls within the Ly$\alpha$ forest.
It is also confused by Ly$\beta$ absorbers 14, 15 and 16
(expected wavelengths indicated by 
short bars at the bottom of the figure).
}
\end{figure*}

The complex CIV absorption feature 
at outflow velocity -2200 -- -3400 kms$^{-1}$
(Fig. 8)
is a `mini-BAL', since
the total velocity range is $<$ 2000 kms$^{-1}$.

Five individual velocity components can be identified, with the CIV
doublet well-resolved in each case.
The velocity components are listed in Table 4.
All are also detected in NV (Fig. 9), but with some blending of
the components because of the lower spectral resolution.
An additional absorber (13) is detected in NV.
The mini-BAL is also detected in OVI (Fig. 10), but is
confused by the Ly$\alpha$ forest.
Absorber 10 is
detected in SiIV (Fig. 6).

The mini-BAL is observationally
unusual in exhibiting significant velocity structure, but
without blending between the two components of the CIV doublet.
This allows the covering factor and optical depth to be measured
independently.

The variation of covering factor with velocity (below)
and the large velocity spread (much greater than 
that expected for a galaxy halo)
suggest that the mini-BAL is intrinsic to the quasar.

\subsubsection{Local covering factor and optical depth}
The observed depth of an absorption feature depends on the 
optical depth $\tau$ of the absorbing cloud, and on 
the fraction $C$ of the
source which the cloud covers
(or, more generally, on $C$ as a function of $\tau$).
The similarity of the CIV and NV mini-BAL profiles
(Figs. 8, 9)
suggests that the form of the mini-BAL is dominated by
variations in local
covering factor $C$ rather than by variations in optical depth.
If the form of the mini-BAL
were due to variation of optical depth, it's unlikely,
given the different ionisation potentials and abundances of CIV and NV,
 that
this would result in similar CIV and NV absorption profiles.

Given observations with sufficient signal-to-noise of
two absorption lines of a particular ion
for which the ratio of the optical depths is known,
one can solve for both $C$ and $\tau$.
For the CIV doublet, the ratio of the optical depths
in the blue and red 
components is that of the oscillator strengths,
$\tau$ (red) = $\tau$ (blue) / 2.
Then the residual intensities in the red and blue
components of the line, expressed as a fraction of the continuum
intensity, are (Barlow \& Sargent 1997, Arav et al 1999):

$I_r = (1-C) + Ce^{-\tau/2}$

$I_b = (1-C) + Ce^{-\tau}$

where $\tau$ is the optical depth in the blue component of the doublet.
These equations can be solved for $C$:

$C = (I_r^2 - 2I_r + 1 ) / (I_b - 2I_r + 1)$ 

(as long as $ I_r \geq I_b \geq I_r^2$) and for $\tau$.

The derivation of $C$ and $\tau$ from $I_b$ and $I_r$ is shown
graphically in Fig. 11.
Permitted combinations of $I_b$ and $I_r$ (i.e. satisfying the above 
inequalities)
fall in the unshaded region of the figure.
The figure highlights the need for good S:N in carrying out
this type of analysis, with the errors on derived $C$ and $e^{-\tau}$
being comparable to the errors on $I_b$, $I_r$ over much of the
allowed range.
It also gives an overview of
the effects of errors in $I_b$ or
$I_r$ due to inadequate spectral resolution or to contamination of one
of the two lines. 
For example, with inadequate resolution, a narrow saturated feature,
with 
$C$ = 1, $e^{-\tau}$ = 0, and true $I_b$ = 0, $I_r$ = 0, may be
observed as a broader unsaturated feature, with $I_b >$ 0, $I_r >$ 0,
causing $C$ to be underestimated, and $e^{-\tau}$ to be 
overestimated.
Fig. 11 also illustrates how $C$ and $\tau$ are constrained by 
mere limits on $I_b$ and $I_r$.
E.g. if $I_b <$ 0.2, $C$ must be $>$ 0.8.

The CIV mini-BAL shows similar velocity structure
in each component of the doublet
(Fig. 8), and the components
are (just) unblended, with the intensity between them, near 
observed wavelength 6724 \AA,
being close to that of the continuum just blueward or redward of the
mini-BAL.
The derived values of $C$
and $\tau$ for the CIV mini-BAL (absorbers 10 - 12)
are shown as a function of velocity in Fig. 12a.
Note that the locus of 
measured $I_b$, $I_r$ on Fig. 11 remains within the
region of physically meaningful
solutions 
$ I_r \geq I_b \geq I_r^2$
(confirming that the spectrum
has sufficient signal-to-noise to solve reliably for $C$, $\tau$).
Fig. 12a may be compared with fig. 2 of Arav et al (1999), who carried
out a similar analysis, of a Keck spectrum of the mini-BAL in 
radio BAL-like quasar 1603+3002.
In 1624+3758, the shape of the absorption in the CIV mini-BAL
also appears to be dominated by variations
with velocity of $C$ rather than variations of $\tau$.
For outflow velocities $<$ -2500 kms$^{-1}$, the optical depth is consistent
(within the errors, $\approx$ 0.03 in $I_b$, $I_r$) 
with $e^{-\tau}$ = 0, i.e. saturated, indicating that
the true optical depth is
much larger than is implied by the depth of the absorption 
feature.
The limited covering factor $C \sim$ 0.6, and the variation of $C$
with velocity,
suggest that the mini-BAL is intrinsic to the quasar.
The mean covering factor for $v <$ -2500 km s$^{-1}$ is $\approx$ 0.7,
that for $v >$ -2500 km s$^{-1}$ is $\approx$ 0.6.

Four effects could complicate the above determination of $C$ and $\tau$
for the CIV mini-BAL.

(1) The covering factor of 
features narrower than the spectral resolution of ISIS
(0.8 \AA, 36 kms$^{-1}$) may be underestimated.
In particular, the observed absorption minima (FWHM 1.3, 0.9 and 1.5
\AA, Fig. 8) might be due to
saturated lines of similar equivalent widths.
Saturation implies $I_r$, $I_b$ = 0, $C$ = 1, $e^{-\tau}$ = 0.
In addition, 
inadequate spectral 
resolution can mimic partial covering, $C <$ 1, in the (instrumental)
wings of deep absorption features
(Ganguly et al 1999).
In the CIV mini-BAL anlaysed here, absorbers 10 and 11 are
separated by five times the instrumental FWHM (0.8 \AA), 
and absorbers 11 and 12
by four times the FWHM, so over most of the wavelength range in
these intervals, there should be negligible 
contamination by the nearby absorption minima.
We confirmed this by convolving with the
instrumental PSF, a simulated spectrum
comprising three saturated features (i.e. $C$ = 1)
at the velocities of absorbers 10, 11 and 12, and with widths
1.0, 0.6 and 0.5 \AA\
(36, 22 and 18 kms$^{-1}$) respectively, to give the same equivalent
widths as observed.
The values of $I_b$ and $I_r$ measured from the convolved spectrum
differ from those in the unconvolved spectrum
by $<$ 0.05, over more than half of the total velocity range
-2300 $< v <$ -2750 kms$^{-1}$.
This confirms that over most of the
resolution elements between the absorption minima, 
 $C$ is smaller than the lower limit on $C$
within the minima, i.e. it supports our conclusion 
that variations of
$C$ dominate the shape of the observed absorption.

(2) The blue component of absorber 10 
is blended with the red component of absorber 9 (Fig. 8), 
so the measured $I_b$ = 0.15 for the former could be underestimated
by up to a few tenths.
However, 
from Fig. 11 it can be seen that given the measured (and unblended) 
$I_r$
= 0.25 in the minimum of
absorber 10, $C$ at this velocity (-2700 $\pm$  50 \AA),
is at most overestimated by $\sim$ 0.15.

(3) It is assumed above that the BAL 
clouds cover both the nucleus (continuum) and the broad-emission-line
region (BLR).
If the BAL covers the source of the continuum, but not the BLR
(as is perhaps the case for the $z$ = 3.3498 absorber
in 1624+3758, Section 3.3, and for
the mini-BAL/NAL 
absorber discussed by Arav et al 1999), then the residual intensities
would have to be measured relative to the continuum rather than
continuum + BLR.
Judging from Fig. 4, the contribution of the BLR at the wavelength 
of the mini-BAL adds $\sol$ 15\% to the light from the continuum, 
and varies little with wavelength across the mini-BAL,
so the measured $I_b$, $I_r$  would have to be revised 
downwards by $<$ 0.15 if the mini-BAL does not
cover the BLR.  This would increase the estimate of the 
covering factor $C$ by $\approx$ 0.15, but leaves the estimate of
$e^{-\tau}$ almost unchanged.

(4) The equations relating $I_b$, $I_r$, $C$ and $\tau$
are based on the assumption of homogeneous partial coverage i.e. an 
opaque screen covering a fraction $C$ of the source.
In reality, as emphasised by Hamann \& Sabra (2004), absorbers are 
probably inhomogeneous, with 
different fractions of the source covered by absorbers of different
opacity.
However, the modelling of Hamann \& Sabra suggests that $e^{-\tau}$ 
as measured above will typically not differ by more than
a few tens \% from the true  
spatially-averaged value.

The residual intensities of the blue and red components of absorber
8, just blueward of the other components of the mini-BAL,
but probably physically associated, imply covering
factor $\approx$ 0.5, and near saturation.
No attempt has been made to model the variation of
covering factor with velocity, because
of blending.

In NV (Fig. 9), the velocity structure is not well-resolved (the spectral resolution
is poorer, $\approx$ 130 kms$^{-1}$),
so that a similar analysis (Fig. 12b) tends to underestimate
the derived covering factor in the minima, but the mean
covering factor $C \sim$ 0.7, is similar to that found
for CIV, and the opacity appears to be less at
outflow velocities -2300 -- -2450 kms$^{-1}$, as for CIV.

A similar analysis in OVI (Figs. 10, 12c, same spectral resolution as for NV) 
is confused by the Ly$\alpha$ forest (and also
by Ly$\beta$ absorbers 15 and 16).
The red component of the OVI doublet due to absorber 10
(which is unconfused by the Ly$\beta$ absorption)
is black, i.e. the covering factor $C$ = 1, and $e^{-\tau}$
= 0,
in contrast to that for CIV and NV ions.
Absorber 11 also appears to have $C \approx$ 1.
Not much can be said about the variation of $C$ and $\tau$
with velocity.
The poor S:N in the OVI mini-BAL 
results in $I_b > I_r$ (non-physical, see Fig. 11) 
over some of the velocity range.
For these velocities, the plotted curves in Fig. 12c
give $C$ = 1 - $I_r$, and
$e^{-\tau}$ = 0.

The variation of $C$ with ion 
(found also by Arav et al 1999, for 1603+3002)
supports the suggestion
by Hamann \& Sabra (2004) that a given observed covering factor $C$
is often due to
inhomogeneous partial coverage, in which optical depth changes 
across the source.
The stronger transitions can then have $\tau >$ 1 over larger areas than
weak transitions.

Approximate 
column densities and limits for the apparently less-saturated part of the
mini-BAL (-2300 -- -2450 km s$^{-1}$, $\tau \sol$ 1),
are given in Table 5.
Column-density measurements of this accuracy do not justify
detailed photoionisation calculations, but
the HI, OVI, NV and CIV 
column densities are consistent with an ionisation parameter
log$(U)$ = -1.5, log$_{}$ $N$(H$_{total}$) $\sim$ 18.4.
using the 
photoionisation modelling of
Hamman (1997, fig. 2c), with 
ionising continuum $F_\nu \propto \nu^{-1.5}$,
and meteoritic solar abundances from Grevesse \& Anders (1989).
The lack of SiIV (and SiII) absorption implies
log$_{} $$U >$ -2.0, i.e. is consistent with the above.

\begin{figure*}
\centering
\psfig{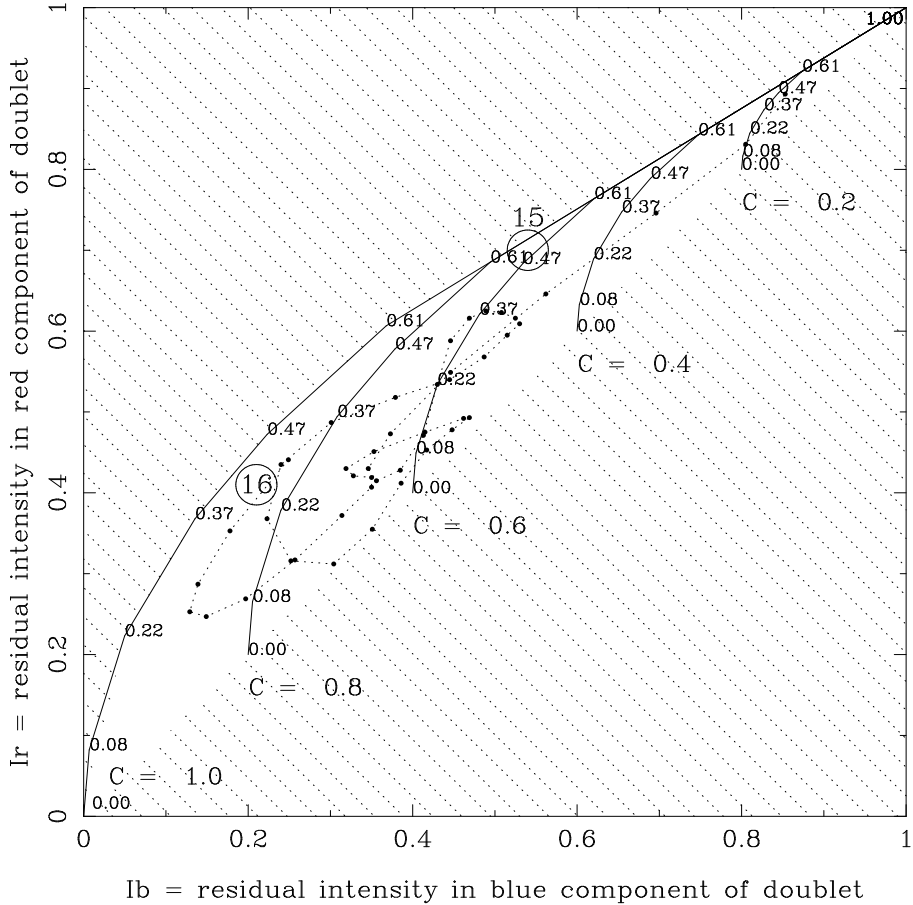}
\caption{Derivation of the covering factor $C$ and optical depth $\tau$ 
from the residual fractional intensities $I_b, I_r$ in the
two components of a doublet with expected optical depth ratio 2:1
(e.g. CIV 1548.2 / 1550.8 \AA, NV 1238.8 / 1242.8 \AA,
OVI 1031.9 / 1037.6 \AA).
Each solid curve traces for a given covering factor $C$,
the expected variation of $I_b$, $I_r$ with  
$e^{-\tau}$. The small numbers on the plot give $e^{-\tau}$ for
the blue component of the doublet.
Combinations of $I_b$, $I_r$ in the shaded region of the figure
are non-physical (see Section 3.2.1).
The dotted curve joins the $I_b$, $I_r$ values (large dots) measured 
for the CIV mini-BAL (Fig. 12a) over the velocity range
-2730 -- -2280 kms$^{-1}$.  The curve lies within the permitted region,
confirming that at most velocities
the solution for $C$, $e^{-\tau}$ is well defined.
The rms measurement errors on $I_b$, $I_r$ are $\sim$ 0.03, and the errors
in derived C and $e^{-\tau}$ will be similar.
The large circles indicate the values (and rms error radius) of
$I_b$, $I_r$ for the 2 associated narrow-line
CIV absorbers 15 and 16, discussed in Section 3.5.
}
\end{figure*}

\subsubsection{Possible line-locking}

The CIV mini-BAL comprises 5 readily-identified CIV doublets 
(absorbers 8 - 12, Fig. 8).
The blue component of the strongest, absorber 10, lies
at 6715.8 \AA, just 0.2 $\pm$ 0.2 A blue of the 6716.0 expected wavelength
of the red component of absorber 9.
This suggests that the lines may be locked together.
The chance of the red component of a given CIV doublet falling within
$d\lambda$ of the blue component of any other CIV doublet, for $N$ doublets
distributed at random over wavelength range $\Delta\lambda$, is 
$d\lambda/\Delta\lambda*N*(N-1)/2$.
For $\Delta\lambda$ = 40 \AA, $d\lambda$ = 0.2, 
$N$ = 5, this probability is 0.05,
i.e. the above coincidence is moderately significant.

The sixth absorber in the mini-BAL (no. 13) was 
identified in NV only (Fig. 9,
expected CIV wavelengths marked on Fig. 8).
Interestingly,
the red component of the CIV doublet of absorber 10 lies at 6727.1 \AA,
0.1 $\pm$ 0.2
\AA\ from the 6727.0 \AA\ expected wavelength of the blue component
of absorber 13.
Absorbers 9, 10 and 13 might therefore be line-locked together
(Fig. 8).

Absorption-absorption line-locking can occur 
when light of the wavelength required for a given transition
in one cloud is absorbed by ions 
in a cloud closer to the quasar, with different velocity and
undergoing
a different transition.
This reduces the radiation force on the shadowed cloud,
and the cloud may 
lock at a velocity difference from the shadowing cloud 
corresponding to the wavelength difference of the two transitions.
In general, several lines will contribute to the total radiation 
pressure on a cloud, but if this approximates the net force in the
opposite direction (gravity, and perhaps drag), the effect of line-locking
in one line could be significant (Korista et al 1993).
Observation of line-locking 
lends support to the hypothesis that radiation pressure 
plays an important
role in the acceleration of BAL gas (in at least some quasars).
Line-locking is probably seen in the spectra of
$z$ = 1.8 quasar 1303+308 (Foltz et al 1987, Vilkovskij \& Irwin 2001)
and $z$ = 2.9 quasar 1511+091 (Srianand et al 2002).
The former includes
several SiIV absorption doublets spaced by the separation
of the two components of the doublet.
Plausible examples of line-locking have also been noted in
08279+5255, 1230+0115, 1303+308, 1511+091, 1605-0112,
NGC 5548 (references given in Section 1.1).


\begin{table*}
\begin{minipage}{170mm}
\begin{center}
\caption{Absorption lines detected in the spectrum of 1624+3758}
\begin{tabular}{rrrlrrrrrrl}
\hline
Ref. &$z_{abs}$ &Velocity& Lines detected& d$\lambda$ & $R$
  &FWHM$_b$&$EW_b$&FWHM$_r$&$EW_r$& Notes \\
no. &    &km s$^{-1}$    &             & \AA\ &\AA\ &\AA\ &\AA&\AA&\AA &  \\
(1) & (2) & (3) & (4) & (5) & (6) & (7) & (8) & (9) & (10) & (11) \\
\hline
 1&2.6379&-54863& Ly$\alpha$ CIV                &        -0.1 & 2.3 
   &2.1 &0.24 &1.7 &0.19 &  \\     
 2&2.7977&-42307& Ly$\alpha$ CIV               &            & 1.8
   &1.6 &0.16 & &$\sim$ 0.15 & Second line of CIV   \\
  &      &     &                  &            &
   & & & & &   doublet obscured \\
  &      &     &                  &            &
   & & & & &   by NaD sky line\\
 3&2.9722&-29022& Ly$\alpha$ SiIV CIV         &       -0.1 & 1.8
   &1.6 &0.14 &$\sim$ 2.5 &$\sim$ 0.21 &  \\ 
 4&3.0438&-23704& Ly$\alpha$ CIV               &        0.2 & 1.8
   &3.1 &0.30 &$\sim$1.6 &0.11 & Resolved \\
 5&3.1182&-18262& Ly$\alpha$ CIV               & $\sim$ 0.5 & 1.8
   &2.2 &0.12 &2.0 &0.05 &  \\
 6&3.2251&-10592& Ly$\beta$ Ly$\alpha$ CIV            & $\sim$ 0.6 & 0.8
   &1.4 &0.11 &$\sim$ 2.6 &0.17 &  Resolved\\
 7&3.2765& -6967& Ly$\beta$ Ly$\alpha$ SiIV CIV         & $\sim$ 0.0 & 0.8
   &2.5 &0.34 &4.1 &0.24 & Resolved \\
\hline
8&3.3283& -3356& OVI NV CIV   & $\sim$ 0.1 & 0.8
   & & & & &  \\
9&3.3307& -3190&CIV               &            & 0.8 
   & & & & &   \\
10&3.3378& -2699& OVI Ly$\alpha$ NV SiIV CIV
                              &   0.0 & 0.8
   & & & & &  \\
11&3.3405&-2512 & OVI Ly$\alpha$ NV CIV&       -0.3 & 0.8
   & & & & &   \\
12&3.3429& -2346& NV CIV      &       -0.1 & 0.8
   & & & & &   \\
13&3.3450& -2201&NV               &             & 0.8
   & & & & &   \\
\hline
14&3.3498& -1870&Ly$\beta$ Ly$\alpha$            &    & 0.8
   & & & & &  Possible LLS \\
15&3.3650& -824& Ly$\beta$ Ly$\alpha$ SiIV CIV            &   0.6 & 0.8
   &2.1 &0.22 &1.4 &0.10 &  Strong  CIV, resolved \\
16&3.3803&  206&Ly$\beta$ Ly$\alpha$ SiIV CIV             &   0.3 & 0.8
   &1.8 &0.32 &2.0 &0.26 &  Strong CIV, resolved \\
\hline
\end{tabular}
\\
\end{center}
The columns give: 
(1) Absorber reference number;
(2) Redshift measured from the blue component of the CIV doublet
    (or NV for absorber 13, Ly$\alpha$ for absorber 14), rms error 0.0001;
(3) Absorber velocity relative to the assumed quasar redshift 3.377
    (Section 3.1), calculated using
    $v/c = (R^2-1)/(R^2+1)$, $R=(1+z_{em})/(1+z_{abs})$;
(4) Ions detected in absorption;
(5) Observed - expected ($\approx$ 11 \AA)
    separation of CIV doublet (observed-frame);
(6) Resolution of the spectrum from which the FWHM and 
    equivalent widths were measured;
(7-10) Observed-frame 
    FWHM and rest-frame equivalent widths of blue and red components of CIV
    absorption line;
(11) Notes.
All of the CIV 
absorbers with $z >$ 3.0 are detected in both the June 18 and July 2
ISIS red-arm
spectra, except that absorber 9 is detected only in the (higher-resolution)
July 2 spectrum.

Absorbers 1-7 are 
 probably intervening absorbers (i.e. not associated with the
quasar).
Absorbers 8 - 13 are components of the mini-BAL.
Absorbers 14 - 16 are probably intrinsic, `associated', absorbers.
\end{minipage}
\end{table*}

\begin{table*}
\begin{minipage}{170mm}
\begin{center}
\caption{Derived column densities of absorbers in 1624+3758}
\begin{tabular}{lrrrrrrr}
\hline
Absorber & Velocity & \multicolumn{5}{c}{log$N$ (cm$^{-2}$)} & Cov \\
(and ref. no.)& kms$^{-1}$ & HI & OVI & NV & SiIV & CIV & fac\\
(1) & (2) & (3) & (4) & (5) & (6) & (7) & (8)\\
\hline
BAL       & -21000 -- -29000 &   &   &&&$>$16.0 &0.3\\
Mini-BAL (12) & -2300 -- -2450     &$>$ 14&15.1&14.7&$\leq$13.2&14.7 
&0.6\\
LLS (14) & -1870 & 18.6&$<$13.6&$<$13.4&$<$12.5&$<$13.1 
&0.7\\
AAL (15) & -824 & $\sol$18.0&&&13.0&13.7 &1.0\\
AAL (16) & 206& $\sol$18.0&&&13.0&13.9 &1.0\\
\hline
\end{tabular}
\\
\end{center}
For the BAL and mini-BAL, 
the column densities or limits were calculated
using $N(v) = 3.77e14 \tau f^{-1} \lambda^{-1} dv$
cm$^{-2}$ (Savage \& Sembach 1991), where 
the oscillator strengths
$f$ are taken from Verner, Barthel \& Tytler (1994),
laboratory wavelengths 
$\lambda$ are given in Table 3, and
$v$ is in km s$^{-1}$.
The metal-line column densities for the LLS and AALs were calculated
from the rest-frame equivalent widths $EW$, using the alternative formulation:
$N$ = $1.13 \times 10^{20} EW f^{-1} \lambda^{-2}$ cm$^{-2}$ (Morton 2003). 
The $N$(HI) column density of the LLS (absorber 14)
is from a Voigt-profile fit to
the damping wings (Fig. 13).
The $N$(HI) column density limits for the AALs are derived from 
the Ly$\alpha$ curve
of growth.
The covering factors $C$ (column 8) are for CIV, 
except for the LLS, $C$ measured from
Ly$\alpha$.
\end{minipage}
\end{table*}

\subsection{Possible Lyman-limit system at $z$ = 3.3498}
\begin{figure}
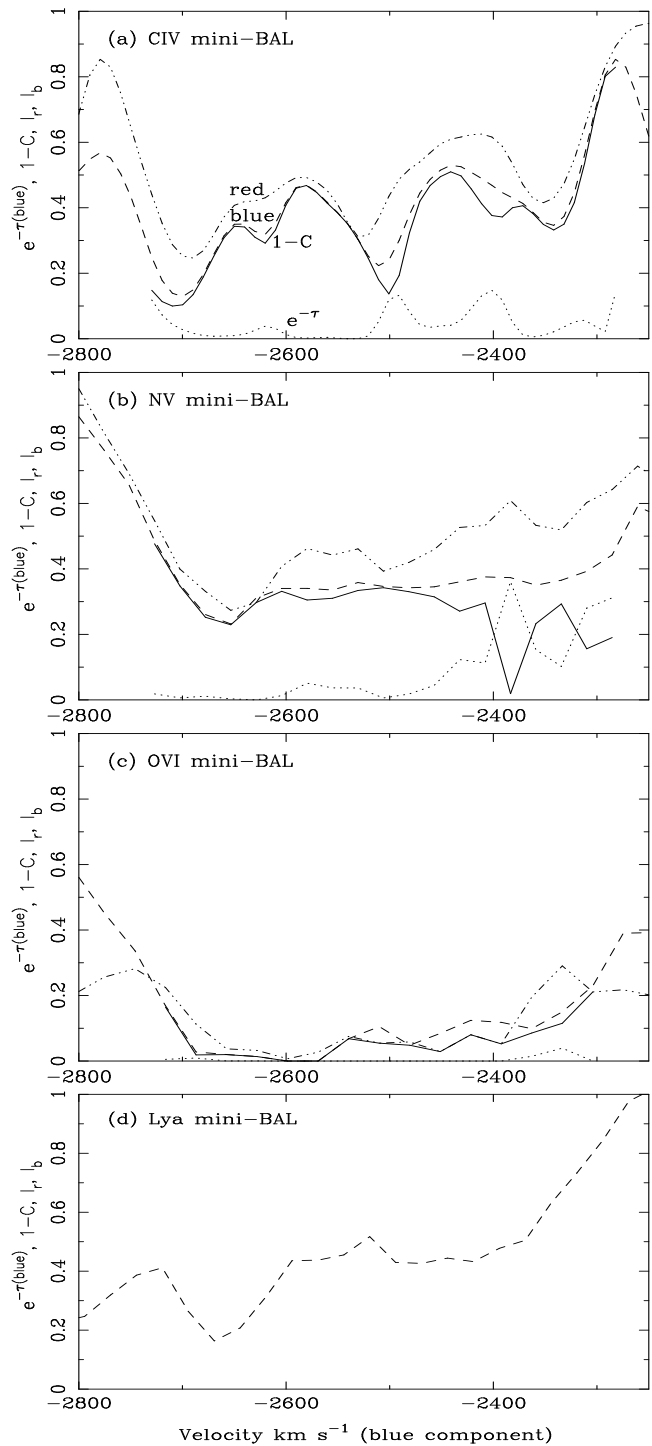

\begin{tabular}{c}
\psfig{file=figcovciv.ps,width=8.5cm,height=4.7cm,angle=270}\\
\psfig{file=figcovnv.ps,width=8.5cm,height=4.7cm,angle=270}\\
\psfig{file=figcovovi.ps,width=8.5cm,height=4.7cm,angle=270}\\
\psfig{file=figcovlya.ps,width=8.5cm,height=4.7cm,angle=270}\\
\end{tabular}
\caption{
Covering factor $C$ (the solid curve shows 1-$C$) 
and optical depth $e^{-\tau}$ 
(in the blue component of the doublet,
dotted) derived as a function of velocity for the mini-BAL
in (a) CIV, (b) NV and (c) OVI.
The residual intensities $I_b$ and $I_r$ from which $C$ 
and  $e^{-\tau}$ were derived are
also shown (dashed and dot-dashed curves `blue' and `red' respectively).
The shape of the CIV mini-BAL is dominated by changes of the
covering factor with velocity, rather than by changes in optical depth.
$C$ may be overestimated near velocity
-2700 kms$^{-1}$ due to blending with absorber 9 (see Section 3.2.1).
(d) The Ly$\alpha$ absorption over the same velocity range.
The spectral resolution for NV, OVI and Ly$\alpha$
(Fig. 12bcd) 
is a factor of 3 poorer than for CIV (Fig. 12a).
}

\end{figure}

The absorption feature at 5288 \AA\ (Figs. 5, 13, 14) 
is flat-bottomed and 
shows possible damping wings, both suggesting saturation.
The feature might be a blend of weaker lines, but here we
explore the possibility that this is a saturated Ly$\alpha$
absorber at  $z$ = 3.3498.
The residual flux at the base of line cannot be due to an 
intensity-calibration problem, since many of the nearby 
Ly$\alpha$-forest lines reach zero intensity.
It therefore implies partial coverage of the source, 
$C$ = 0.7.
The rest-frame equivalent width of the line is 1.5 \AA.
A Voigt-profile fit (Fig. 13) yields
column density $N$(HI) = 4$\times$10$^{18}$ cm$^{-2}$,
velocity parameter $b$ = 30 km\,s$^{-1}$, 
i.e. a Lyman-limit system
(LLS, 17.2 $<$ log$N$(HI) $<$
20.3, e.g. Lanzetta et al 1995).

The corresponding
Ly$\beta$ falls in a heavily-absorbed part of the Ly$\alpha$ forest,
but may be detected, with the expected rest-frame equivalent 
width $\sim$ 1.0 \AA.
The Lyman limit would be at 3967 \AA, but
the SDSS spectrum has zero intensity bluewards of 3980 \AA,
perhaps due to Lyman-limit absorption by absorber 15 (see Section 3.5).
Surprisingly, no metal lines are detected, $N$(SiII), $N$(SiIV), 
$N$(CIV) $\sol$ 10$^{13}$ cm$^{-2}$.
This cannot be due to very high ionisation, because no combination 
(Hamann 1997) of
ionisation parameter log$U >$ 0 and $N$(H) $<$ 10$^{24}$ cm$^{-2}$
(above which the gas would be Thompson-thick) is consistent
with the observed limits on metal column density.
It might be due to low metallicity, which
is one possible interpretation of the lack of metal lines 
in another non-black absorber: that
found by Petitjean \& Srianand (1999) in $z$ = 2.2 quasar J2233-606,
with
log$N$(HI) = 14, $C$= 0.7.

The limited covering factor, $C <$ 1, implies that the 
putative absorber is intrinsic,
and close to the nucleus of the quasar.  The  
mini-BAL (Section 3.2) has similar covering factor, 
which might indicate similar physical location.
Alternatively, the absorber might be
covering the quasar continuum source only, 
and not the broad-line region (Fig. 14).

\begin{figure}
\centering
\psfig{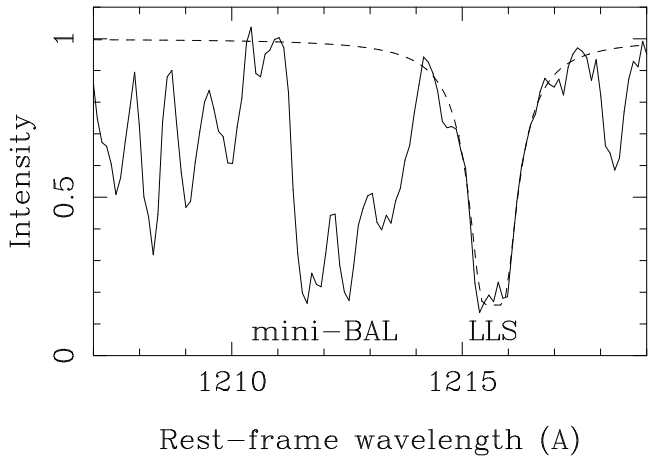}
\caption{
Voigt-profile fit to the $z$ = 3.3498 absorber, $N$(HI) = 4 $\times$ 
10$^{18}$ cm$^{-2}$.
No associated metal lines are detected.
}
\end{figure}

\begin{figure}
\centering
\psfig{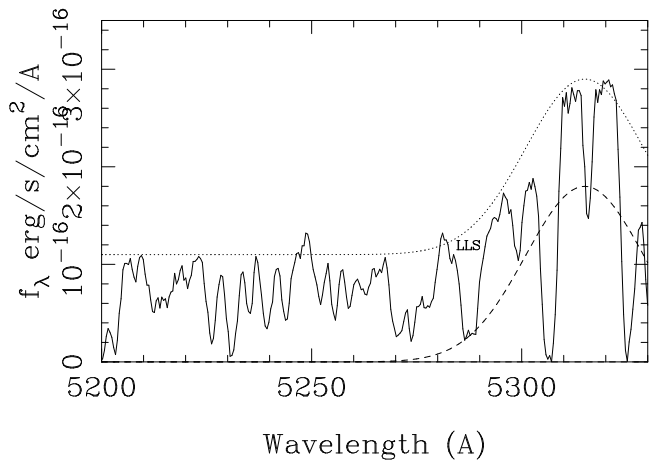}
\caption{
The spectrum of 1624+3758 in the vicinity of 
the $z$ = 3.3498 LLS absorber, showing a possible explanation
of the non-black trough.  The dotted line is an approximate upper
bound to the spectrum blueward of the Ly$\alpha$ emission, 
and is the sum of a wavelength-independent term 
(continuum) and
a gaussian of FWHM = 1700 km s$^{-1}$ (dashed line,
representing broad Ly$\alpha$ emission).
The depth of the $z$ = 3.3498 absorption is consistent with
the cloud covering the continuum
source, but not the broad-line region.
}
\end{figure}

\subsection{BAL (2.968 $< z <$ 3.085)}
The CIV BAL extends over observed wavelengths 6150 - 6330 \AA, 
outflow velocity -29300 -- -20700 kms$^{-1}$.
It is also detected in NV and SiIV, and perhaps 
in Ly$\alpha$ (see Figs. 4, 5),
although the SiIV trough is partially masked by the broad
Ly$\alpha$ and OI emission.
The BALnicity index of 1624+3758 is 
2990 kms$^{-1}$.
The BAL turns on and off over $\approx$ 700 kms$^{-1}$.
The feature bisecting the BAL at 6232 \AA\ corresponds to no known 
emission line, and is probably just a gap in the velocity structure.
The mean depth of the BAL is 0.35 times the intensity of the continuum.
BALs are usually saturated (Hamann, Korista \&
Morris 1993), so the observed depth implies a 
covering factor of 0.35.
The observed spectrum
is consistent with similar covering factor in NV and SiIV.

The BAL is not detected in absorption in the AlIII (1855, 1863 \AA) doublet,
expected wavelengths 7373 - 7590 \AA, (Fig. 4), suggesting that the
quasar is a HiBAL (but see Section 4).

\subsection{Narrow CIV absorption features (NALs)}
In addition to the absorbers discussed above, 9 CIV NALs
are detected (Table 4).  

The CIV lines of
absorbers 15 and 16 are deep, and have small velocities
relative to the quasar ($>$ -1000 kms$^{-1}$).  They are thus likely to
be physically
associated with the quasar.  
Absorber 16 is slightly redshifted
relative to the emission-line redshift of the quasar,
but by only one standard deviation.
Both absorbers are resolved in velocity.
The residual intensities in the blue and red components are 
consistent with covering factor $C$ = 1
(Fig. 11), and moderate optical depth. 
The Ly$\alpha$ lines of both absorbers are saturated, with rest-frame 
equivalent width $\approx$ 0.8 \AA, implying 
$N$(HI) $<$ 10$^{18}$ cm$^{-2}$.
The Ly$\beta$ lines are probably detected (Fig. 10), but are confused
by the OVI mini-BAL.
The Lyman limits for these two absorbers would fall at wavelengths
3981 and 3995 \AA.
The SDSS spectrum shows zero intensity below 3980 \AA,
so absorber 15 may be a Lyman-limit system (LLS,
log$_{}$ $N$(HI) $>$ 17.2).  

The remaining narrow CIV lines (1 -- 7), 
with $v <$ -5000 km/s, 
are likely to be intervening absorbers, physically unrelated to the quasar.
Misawa et al (2002) measured the number density per unit
redshift, $n(z)$, of intervening CIV absorbers.
Combining their results with those of Steidel (1990), 
the mean $n(z)$ = 2 $\pm$ 1 at $z \sim$ 3, for
rest-frame equivalent width $EW >$ 0.15 \AA\ in both
components of the doublet.
For 1624+3758, we find three CIV doublets satisfying these criteria
(absorbers 1, 2, 7)
over a redshift range 0.7, i.e. 4.3 $\pm$ 2 per unit
redshift, not significantly different from that found by Misawa et al,
and Steidel
(and contrasting with the much higher $n(z)$ = 7
found for 0747+2739 Richards et al 2002a).
Of the absorbers noted here, 
4, 6 and 7 are slightly resolved in velocity (Table 3),
consistent with the 
internal velocity dispersions typical
of normal galaxies, $\sim$ 50 kms$^{-1}$.
The rms errors on the equivalent widths in the blue and red components
of each doublet (absorbers 1 -- 7) are large, but the
measurements are consistent with covering factor $C$ = 1,
as expected
for absorbers whose physical dimensions greatly exceed those of
the emitting region.


Most of the NALs are detected in Ly$\alpha$, and several
in Ly$\beta$ (Figs. 5, 10).
None of the NALs are detected in absorption by
OI (1302.17 \AA),
SiII (1526.71 \AA),
FeII (1608.45 \AA),
AlII (1670.79 \AA) or
AlIII (1854.72 \AA).

\section{The nature of 1624+3758}
The lack of 
detectable BAL absorption in the AlIII doublet 
(Section 3.4) suggests that the
quasar is a HiBAL, but the observed spectrum does 
not extend to MgII 2800 \AA, 
so it could also be an atypical LoBAL
with weak AlIII absorption.
The OI equivalent width, and the presence of 
FeII/FeIII emission (Section 3.1)
are typical of LoBALs, as is the continuum colour (although this also
falls within the range of colours of HiBALs).
The frequency of LoBALs amongst radio BAL quasars appears to be higher
$\sim$ 30\%
(Becker et al 2000, 2001,
Menou et al 2001), than amongst non-radio BAL quasars, $\sim$ 15\%.

Regardless of whether the quasar is a HiBAL or a LoBAL, 
the FeII UV191 triplet emission line at 1787 \AA\
(and FeIII at 1896/1914 \AA) is unusually strong (Section 3.1).
The association of unusually high radio luminosity (Section 1.2)
and unusually
strong FeII emission suggests a connection between the two.
Such a connection is also suggested
by the fact that the 1787-\AA\ line is detectable in
only 1 out of 40 SDSS BAL quasars with no FIRST counterparts, but in 
4 out of 14 of the SDSS BAL quasars with $S_{1.4GHz} >$ 1 mJy
(Section 3.1).
Boroson (2002) and Lamy \& Hutsemekers (2004) note that 
strong FeII emission might be a signature of the thickening of the
accretion disk at accretion
rates close to the Eddington limit, $L/L_{edd} \sim$ 1.
Although the intensity of this `small blue bump' FeII emission might not
always be proportional to that of the FeII 1787 \AA\ line
(Vestergaard \& Wilkes 2001), it's plausible that
the unusually strong FeII 1787-\AA\ emission in 1624+3758 is related to 
the very high accretion rates posited for radio BAL quasars
(Boroson 2002).


The detachment of the BAL by 21000 kms$^{-1}$ from the CIV emission
line is moderately unusual, being observed in $\sim$ 10\% of BAL 
quasars with either $z <$ 3 or $z >$ 3.  
It suggests an angle of view well away from
the plane of the accretion disk, so that the line 
of sight to the quasar nucleus
exits the curving streamlines far above the disk
(see fig. 8 of Lamy \& Hutsemekers 2004).

The radio rotation measure of 1624+3758 is the second-largest known.
It is due to the properties of gas lying between
us and the radio-emitting region (size $\sog$
1 kpc), and is unlikely to depend strongly on orientation.
It implies a high value of at least one of 
the magnetic field, electron density or path length 
through the region responsible.

In summary, the observed properties of the quasar are more consistent
with it being intrinsically unusual than with it being viewed at an
unusual orientation.  1624+3758 is highly
radio-luminous, and it may be a good example of
an object which is accreting both at a very high rate (high $dM/dt$) 
and near the Eddington limit.


\section{Conclusions}
We report high-resolution spectroscopy and radio observations
of the BAL quasar 1624+3758, $z$ = 3.377.
1624+3758 is the most
radio-luminous BAL quasar known, 
$P_{1.4GHz}$ = 4.3$\times$10$^{27}$ WHz$^{-1}$.
It is also highly luminous in the optical, $M_{AB}$(1450-\AA) 
$\approx$ -27.6, luminosity $L_{1450}$ 
$\sim$ 5$\times$10$^{24}$ WHz$^{-1}$.

(1) The FeII UV191 1787-\AA\ emission triplet is unusually prominent,
rest-frame $EW$ = 3.8 \AA.
FeIII UV34 1896/1914 \AA\ is also detected.

(2) The BAL has BALnicity index $BI$ = 2990 kms$^{-1}$,
outflow velocity 
-21000 -- -29000 kms$^{-1}$.
The large detachment velocity suggests an angle of view well away from
the plane of the accretion disk.

(3) A complex mini-BAL is detected in CIV, NV and OVI,
with velocity -2200 -- -3400 kms$^{-1}$.
For CIV, we have measured the covering factor and optical depth
as a function of velocity -2300 -- -2700 kms$^{-1}$.
The shape of the absorption is dominated
by the variation with velocity of covering factor.
This variation implies that the mini-BAL is intrinsic to the quasar.

(4) There is statistical evidence of line-locking between 2 
(and perhaps 3) of
the mini-BAL absorption components, 
supporting the hypothesis that the outflows
are accelerated by radiation pressure.

(5) 
A possible non-black HI absorber is observed 
with velocity -1870 kms$^{-1}$, 
$N$(HI) = 4 $\times$ 10$^{18}$ cm$^{-2}$ (LLS).
There are no associated metal lines.
The covering factor is only 0.7,
suggesting that the absorber 
is intrinsic to the quasar, perhaps
covering the continuum source, but not the broad-line region.

(6) The velocities relative to the quasar of two of the CIV NALs 
are small (-824, 206 kms$^{-1}$).
They are likely to be intrinsic.
The other 7 NALs ($v <$ -5000 kms$^{-1}$) are probably intervening systems.
The number density of CIV absorbers with $v <$ -5000 kms$^{-1}$, and
rest-frame equivalent width
$EW >$ 0.15 \AA, is $n(z)$ = 4.3 $\pm$ 2, consistent with  that measured
for other quasars.

(7) The wings of the CIV and NV emission lines are markedly asymmetric, 
consistent with the red wings being suppressed, perhaps due to 
dust extinction of light emitted by gas outflows
on the far side of the nucleus.

(8) The width of the CIV emission line, in conjunction with the optical
  luminosity $L$, implies a black-hole mass $M_{BH} \sim$ 10$^9$ $M_\odot$,
and $L/L_{Eddington} \sim$ 2.

(9)
The radio spectrum turns over at rest-frame $\sim$ 2 GHz,
suggestive of a young compact source.
The source is slightly resolved by the VLA observation, projected size
$\sim$ 2.8 kpc.

(10)
The radio
source is 11\% polarised at 10 GHz, and the rest-frame rotation measure,
18350 rad m$^{-2}$ is the second-highest known for any extragalactic source.

(11) The conjunction of several unusual features, particularly the
strong FeII 1787-\AA\ emission, 
and the high radio rotation measure, 
favour the quasar being intrinsically unusual rather than
being viewed at an unusual orientation.
Given the high radio luminosity,
the unusual features may be due to a combination of a 
very high accretion rate and high $L/L_{Eddington}$,
i.e. this quasar may occupy an extreme position in Boroson's (2002)
classification scheme for AGN.



{\bf Acknowledgments}
\\We are grateful to Pierre Leisy for obtaining one of the spectra 
during WHT service time,
to Marek Jamrozy for making the Effelsberg observations, and
to the anonymous referee for helpful suggestions.
CRB, RC, MV and JIGS acknowledge financial support from the
Spanish Ministerio de Ciencia y Tecnolog\'\i a under project
AYA2002-03326.
The William Herschel Telescope is operated on the island of La Palma
by the Isaac Newton Group in the Spanish Observatorio del Roque de
Los Muchachos of the Instituto de Astrofisica de Canarias.
The 100-m Effelsberg radio telescope is operated by the Max Planck Institut
f\"{u}r Radioastronomie.
The National Radio Astronomy Observatory is operated by Associated
Universities, Inc., under cooperative agreement with the National Science
Foundation.
The Westerbork Synthesis Radio Telescope (WSRT) is operated by the Netherlands
Foundation for Research in Astronomy (ASTRON) with financial support of the
Netherlands Organization for Scientific Research (NWO).
The Two Micron All Sky Survey
is a joint project of the University of Massachusetts and the Infrared
Processing and Analysis Center / California Institute of Technology, funded by
NASA and the NSF.
The Sloan Digital Sky Survey is funded by the Alfred P. Sloan foundation,
the SDSS member institutions, NASA, NSF, the US Dept. Energy, 
the Japanese Monbukagakusho and the Max Planck Society.

\end{document}